\documentclass[fleqn,usenatbib]{mnras}

\usepackage{newtxtext,newtxmath}
\usepackage[T1]{fontenc}
\usepackage{ae,aecompl}

\usepackage[utf8]{inputenc}
\usepackage{amsmath}
\usepackage{graphicx}
\usepackage{amsfonts}
\usepackage{verbatim} 
\usepackage{caption}
\usepackage{subcaption}
\usepackage{ulem}

\newcommand{\rhalf}{R_{1/2}}
\newcommand{\rvir}{R_{\rm vir}}
\newcommand{\msun}{{\rm M}_\odot}
\newcommand{\kpc}{{\rm kpc}}
\newcommand{\logit}{{\rm logit}\,}

\title[Galaxy--halo size relation in FIRE]{The galaxy--halo size relation of low--mass galaxies in FIRE}
\author[Rohr et al.]{Eric Rohr$^{1}$\thanks{Contact e-mail: \href{mailto:rohr@mpia.de}{rohr@mpia.de}},
Robert Feldmann$^{2}$,
James S. Bullock$^{3}$,
Onur \c{C}atmabacak$^{2}$,
Michael Boylan-Kolchin$^{4}$, \newauthor
Claude-Andr\'e Faucher-Gigu\`ere$^{5}$,
Du\v{s}an Kere\v{s}$^{6}$,
Lichen Liang$^{7,2}$,
Jorge Moreno$^{8}$,
Andrew Wetzel$^{9}$
\\
$^{1}$Max-Planck-Institut f{\"u}r Astronomie, K{\"o}nigstuhl 17, D-69117 Heidelberg, Germany\\
$^{2}$Institute for Computational Science, University of Zurich, Winterthurerstrasse 190, Zurich CH-8057, Switzerland\\
$^{3}$Department of Physics and Astronomy, 4129 Reines Hall, University of California, Irvine, CA 92697, USA\\
$^4$Department of Astronomy, The University of Texas at Austin, 2515 Speedway, Stop C1400, Austin, TX 78712, USA\\
$^5$Department of Physics and Astronomy and CIERA, Northwestern University, 2145 Sheridan Road, Evanston, IL 60208, USA\\
$^6$Department of Physics, Center for Astrophysics and Space Sciences, University of California, San Diego, 9500 Gilman Drive, La Jolla, CA 92093, USA\\
$^7$Canadian Institute for Theoretical Astrophysics, 60 St. George Street, Toronto, ON M5S 3H8, Canada \\
$^8$Department of Physics and Astronomy, Pomona College, Claremont, CA 91711, USA\\
$^{9}$Department of Physics and Astronomy, University of California, Davis, CA 95616, USA\\
}

\date{Accepted XXX. Received YYY; in original form ZZZ}

\pubyear{2022}

\begin{document}
\label{firstpage}
\pagerange{\pageref{firstpage}--\pageref{lastpage}}
 \maketitle

\begin{abstract}

Galaxy sizes correlate closely with the sizes of their parent dark matter haloes, suggesting a link between halo formation and galaxy growth. However, the precise nature of this relation and its scatter remains to be understood fully, especially for low-mass galaxies. We analyse the galaxy--halo size relation for low-mass ($M_\star \sim 10^{7-9}\, \msun$) central galaxies over the past 12.5 billion years with the help of cosmological volume simulations (FIREbox) from the Feedback in Realistic Environments (FIRE) project. We find a nearly linear relationship between the half-stellar mass galaxy size $\rhalf$ and the parent dark matter halo virial radius $\rvir$. This relation evolves only weakly since redshift $z = 5$: $\rhalf\, [\kpc] = (0.053\pm0.002)(\rvir/35\, \kpc)^{0.934\pm0.054}$, with a nearly constant scatter $\langle \sigma \rangle = 0.084\, [{\rm dex}]$. Whilst this ratio is similar to what is expected from models where galaxy disc sizes are set by halo angular momentum, the low-mass galaxies in our sample are not angular momentum supported, with stellar rotational to circular velocity ratios $v_{\rm rot} / v_{\rm circ} \sim 0.15$. Introducing redshift as another parameter to the GHSR does not decrease the scatter. Furthermore, this scatter does not correlate with any of the halo properties we investigate -- including spin and concentration -- suggesting that baryonic processes and feedback physics are instead critical in setting the scatter in the galaxy--halo size relation. Given the relatively small scatter and the weak dependence of the galaxy--halo size relation on redshift and halo properties for these low-mass central galaxies, we propose using galaxy sizes as an independent method from stellar masses to infer halo masses.
\end{abstract}


\section{Introduction} \label{sec:intro}

According to the standard picture of galaxy formation, galaxies form at the centres of their parent dark matter haloes \citep{White1978,Cole2000}. Haloes tend to form bottom-up, with small haloes collapsing first and subsequently merging into larger and more massive haloes \citep{Blumenthal1984}. This gravity-driven hierarchical picture of halo collapse, however, contrasts with a much more complex picture of galaxy formation which is set by a variety of baryonic processes \citep[e.g.,][and references therein]{Somerville2014}. The link between galaxies and their parent haloes is therefore far from trivial, especially given that the latter are more than order of magnitude larger and more massive than the former.

Nonetheless, central galaxies and their parent haloes do appear to be linked tightly, at least in the local Universe \citep[see][for a recent review]{Wechsler2018}. For instance, the stellar-to-halo-mass relation (SHMR), also called the $M_\star-M_{\rm halo}$ relation, has been widely studied empirically \citep[via abundance matching,][]{Kravtsov1999,Kravtsov2004,Conroy2009,Behroozi2010}, observationally via -- e.g., gravitational lensing \citep{Mandelbaum2006,Behroozi2019} or galaxy kinematics \citep{More2011,Kravtsov2018} -- and theoretically, with the help of analytic models \citep[e.g.,][]{White1991,Mo1998,Wechsler1998,White2007a} and cosmological simulations \citep{Pearce2001,Berlind2003,Simha2012,Hopkins2014,Khandai2015,McAlpine2016,Feldmann2016,Chaves-Montero2016,Pillepich2018a,Engler2020}. This relation forms a broken power law of increasing galaxy formation efficiency with halo mass until a peak at $M_{\rm halo} \sim 10^{12}\, \msun$ and a sharp decrease towards higher halo masses \citep{Wechsler2018}. Interestingly this relation has a nearly constant scatter over five orders of magnitude in halo mass ($M_{\rm halo} \sim 10^{10-15}\, \msun$), and the baryonic feedback processes -- such as stellar winds, supernovae, cosmic rays and active galactic nuclei (AGN) -- largely determine the shape of this relation despite the haloes dominating in size and mass \citep[e.g.,][]{Dekel1986,Silk1998,Bullock2000,Croton2006,Hopkins2012}.

More recently, \citet{Kravtsov2013} discovered a tight connection between the sizes of galaxies and haloes at $z=0$. Unlike the SHMR, this galaxy--halo size relation (GHSR) forms an approximately linear relation with constant scatter across nearly three orders of magnitude in halo size ($R_{\rm halo} \sim 5-1,500\, \kpc$, corresponding to $M_{\rm halo} \sim 10^{8-15}\, \msun$). Furthermore, this relationship is largely independent of galaxy morphology and nearly identical for centrals and satellites. The observation of the GHSR at $z=0$ thus raises a variety of intriguing questions such as: When was the GHSR established? How do its slope, normalization, and scatter evolve over cosmic time? Are the GHSR and its scatter linked to the growth histories of structure in the Universe?

However, observing this galaxy--halo connection at high redshift poses a number of challenges. For instance, measuring galaxy and halo sizes with sufficiently high spatial resolution becomes increasingly more challenging at higher redshift. Fortunately, a variety of approaches have been developed to probe the galaxy--halo link at earlier cosmic times. For instance, \citet{DiTeodoro2015} show that fitting a 3D-tilted ring models can resolve intrinsic rotation curves and velocity dispersion from low spatial resolution observations. Other recent works have recovered connections between galaxies and their host haloes at high redshifts \citep{Shibuya2015,Huang2017,Hirtenstein2019,Zanisi2020,Zanisi2021a}. But the current observations are limited to the most massive galaxies, leaving the low-mass regime uncertain.

Another approach is to simulate galaxy formation in a cosmological context to study the galaxy--halo connection across cosmic history \citep[see][for a recent review]{Somerville2014}. Over the past decade, advances in computational power and numerical techniques \citep[e.g.,][]{Springel2005,Springel2010,Hopkins2015} have made it possible to simulate not only the gravitational collapse of dark matter, but also the hydrodynamics of gas, and the complex baryonic processes such as gas cooling, star formation, and stellar feedback  \citep[e.g.,][]{Vogelsberger2014,Schaye2015,Crain2015a,Tremmel2017,Hopkins2018,Pillepich2018,Nelson2019}. Recent simulations provide a deeper understanding of, and challenge long-standing assumptions about, galaxy formation and evolution. Perhaps contrary to the expectation based on specific angular momentum conservation \citep{Fall1980,Mo1998}, \citet{Desmond2017} find in the EAGLE simulation that at a fixed stellar mass, the galaxy size weakly correlates with halo mass, concentration or spin. On top of that, \citet{Somerville2018} conclude from their sample of Galaxy And Mass Assembly (GAMA) and the Cosmic Assembly Near Infrared Deep Extragalactic Legacy (CANDELS) surveys mapped to the Bolshoi-Planck dissipationless $N-$body simulation that the ratio of galaxy to halo size decreases slightly with cosmic time for less massive galaxies, while the ratio of galaxy size to halo size times halo spin -- $\rhalf / (\rvir \lambda)$ -- is lower for more massive galaxies below $z \lesssim 3$. In the VELA \citep{Ceverino2014,Zolotov2015} and NIHAO \citep{Wang2015} zoom-in simulations, \citet{Jiang2019} find that the halo spin only weakly correlates with that of the galaxy, and this correlation becomes weaker with increasing redshift. Moreover, the gas that builds galaxies in cosmological simulations typically has higher specific angular momentum than that of the dark matter \citep[e.g.,][]{Danovich2015,Stewart2017,Zjupa2017,El-Badry2018,Kretschmer2020}. 

In general, these recent results point to differences between halo and galactic properties once thought to be tightly linked. It thus appears that baryonic properties are more significant in setting galaxy sizes. Using zooms of Milky-Way mass objects with the same FIRE-2 model, \citet{Garrison-Kimmel2018} find that of their studied parameters, the best predictor of galaxy size and morphology is the gas spin at the time the galaxy formed half of its $z=0$ stars. Similarly for massive galaxies ($M_\star \sim 10^{9-12.5}\, \msun$) in the Sloan Digital Sky Survey Data Release 7, \citet{Zanisi2020} find in their semi-empirical models that the specific stellar angular momentum is the best mediator to the GHSR. Despite the numerous works on potential correlations between halo properties and galaxy sizes \citep{Rodriguez2021}, the GHSR has not been well studied in simulations and observations, in particular, in the regime of low stellar masses and at higher redshifts.

In this paper, we study the GHSR and its scatter for low-mass centrals, as predicted by high resolution cosmological volume simulations from the FIREbox simulation suite, which is part of the \textit{Feedback in Realistic Environments} (FIRE)\footnote{\url{https://fire.northwestern.edu/}} project. FIREbox follows the growth of galaxies and haloes in a $15\, {\rm cMpc}\, h^{-1}\ (22\, \text{pMpc at }z=0)$ side-length cosmological box with the help of the FIRE-2 baryonic model \citep{Hopkins2018} and the Meshless Finite Mass hydrodynamic solver {\sc gizmo}\footnote{\url{http://www.tapir.caltech.edu/~phopkins/Site/GIZMO.html}} \citep{Hopkins2015}. This simulation suite contains a large sample of galaxies ranging from isolated dwarfs to Milky-Way (MW) analogs, facilitating an in-depth analysis of the GHSR from $z=5$ until today.

The layout of this paper is as follows. In \S~\ref{sec:meth} we outline the methodologies for the numerics of the simulation (\S~\ref{sec:NR}), the halo finding algorithms (\S~\ref{sec:HF}), and the sample selection (\S~\ref{sec:SS}). \S~\ref{sec:comp} compares the galaxy/halo pairs to observations and other recent works. \S~\ref{sec:GHSR} constructs and details the GHSR from $0 \leq z \leq 5$. \S~\ref{sec:scat} analyzes the scatter in the GHSR and addresses potential halo (\S~\ref{sec:haloprop}), galaxy (\S~\ref{sec:galprop}), and environment (\S~\ref{sec:envirprop}) properties affecting the $\rhalf-\rvir$ relation. Specifically, \S~\ref{sec:halocon_spin} investigates the effects of halo spin and concentration on the GHSR and SHMR in more detail. Lastly, we summarise the major findings in \S~\ref{sec:con}.

\section{Methodology} \label{sec:meth}

\subsection{FIREbox simulation suite} \label{sec:NR}

The galaxies and haloes analysed in this paper are extracted from the FIREbox suite of $V=(15\,{\rm cMpc}\,h^{-1})^3$ cosmological volume simulations (Feldmann et al. in prep), which are part of the FIRE project \citep{Hopkins2014,Hopkins2018}. Unlike all previous FIRE simulations, FIREbox does not use the zoom-in set-up to study galaxy evolution -- but instead, it simulates gas, stars, and dark matter in a cubic cosmological volume with periodic boundary conditions. Initial conditions at $z=120$ were created with MUlti Scale Initial Conditions \citep[MUSIC;][]{Hahn2011} using cosmological parameters consistent with Planck 2015 results \citep{Planck2015}: $\Omega_{\rm m}=0.3089$, $\Omega_\Lambda=1-\Omega_{\rm m}$, $\Omega_{\rm b}=0.0486$, $h=0.6774$, $\sigma_8=0.8159$, $n_{\rm s}=0.9667$ and a transfer function calculated with camb\footnote{\url{http://camb.info}} \citep{Lewis2000,Lewis2011}.

All FIREbox simulations start from the same initial conditions but they differ in particle number, numerical resolution, and whether they are run as dark-matter-only (DMO) simulations or include baryonic physics. All simulations are run with \textsc{gizmo} \citep{Hopkins2015}. Gravitational forces between particles are calculated with a heavily modified version of the parallelisation and tree gravity solver of GADGET-3 \citep{Springel2005} allowing for adaptive force softening, while hydrodynamics is solved with the meshless-finite-mass method introduced in \citet{Hopkins2015}. All hydrodynamical FIREbox simulations are run with the FIRE-2 model to account for gas cooling and heating, star formation, and stellar feedback \citep{Hopkins2018}. Feedback from supermassive black holes is not included. Star formation is modeled to occur in dense ($n>300$ cm$^{-3}$ for the 1024$^3$ FIREbox simulation; $n > 100$ cm$^{-3}$ and $>10$ cm$^{-3}$ for the 512$^3$ and 256$^3$ simulations respectively), molecular, self-gravitating gas, and the gas to star conversion takes place on a local free-fall time with a 100\% local efficiency.  Due to stellar feedback, the realized star formation efficiency is lower, consistent with Kennicutt-Schmidt relations \citep{Schmidt1959,KennicuttJr.1998,Orr2018}. Stellar feedback includes energy, momentum, mass, and metal injections from supernovae (type II and type Ia) and stellar winds (OB and AGB stars). Radiative feedback (photo-ionisation and photo-electric heating) and radiation pressure from young stars is accounted for in the locally extincted background radiation in optically thin networks (LEBRON) approximation \citep{Hopkins2012a}. The FIRE-2 model has been extensively validated in a number of publications analysing properties of galaxies across a range in stellar masses and numerical resolutions, including simulations at this FIREbox resolution \citep{Wetzel2016,Hopkins2018,Ma2018a,Ma2018}. 

Most of the analysis in this paper is based on the FIREbox pathfinder hydrodynamical simulation ($N_{\rm b}=1024^3$ and $N_{\rm DM}=1024^3$). The mass resolution of this run is $m_{\rm b}=6.3\times{}10^4$ $M_\odot$ for baryonic (gas and star) particles and $m_{\rm DM}=3.3\times{}10^5$ $M_\odot$ for dark matter particles. The force softening lengths for star and dark matter particles are $h_{\star}=12\, {\rm pc}$ (physical) and $h_{\rm DM}=80\, {\rm pc}$ respectively. The force softening of gas particles is set to their smoothing length down to a minimum of 1.5 pc, which is reached only in the densest parts of the interstellar medium. The force resolution is set such that the highest density we formally resolve is 1,000 times the star formation threshold \citep[see][\S~2.2 for more details]{Hopkins2018}. Mass and force resolution of the $N_{\rm DM}=512^3$ ($N_{\rm DM}=256^3$) FIREbox run are correspondingly lower, e.g., $m_{\rm b}\sim{}5\times{}10^5$ $M_\odot$ ($m_{\rm b}\sim{}4\times{}10^6$ $M_\odot$) and $h_{\rm star}=32$ pc ($h_{\rm star}=128$ pc). All FIREbox simulations examined in this paper are evolved to $z=0$.

\subsection{Halo Finding and Definitions} \label{sec:HF}

We employ the AMIGA Halo Finder (AHF)\footnote{\url{http://popia.ft.uam.es/AHF/Download.html}} to identify and characterise the properties of dark matter haloes \citep{Knollmann2009}. We only consider haloes containing at least $100$ particles of any type, which corresponds to a minimum halo mass of $M_{\rm vir}\sim10^7\, \msun\, h^{-1}$. The halo radius, $\rvir$, is defined based on the virial overdensity criterion
\begin{equation} \label{eqn:mvir}
M_{\rm vir} = (4/3) \pi \Delta(z) \rho_{m}(z) \rvir^3,
\end{equation}
where $\rho_{m}(z)$ is the matter density at a given redshift $z$, and $\Delta(z)$ is defined by \citet{Bryan1998}. Halo centres and the centres of their central galaxies are identified as the halo region with the highest total matter density using AHF's maximum-density (MAX) setting.

\subsection{Sample Selection and Galaxy Definitions} \label{sec:SS} 

\begin{figure*}
    \includegraphics[width=\textwidth]{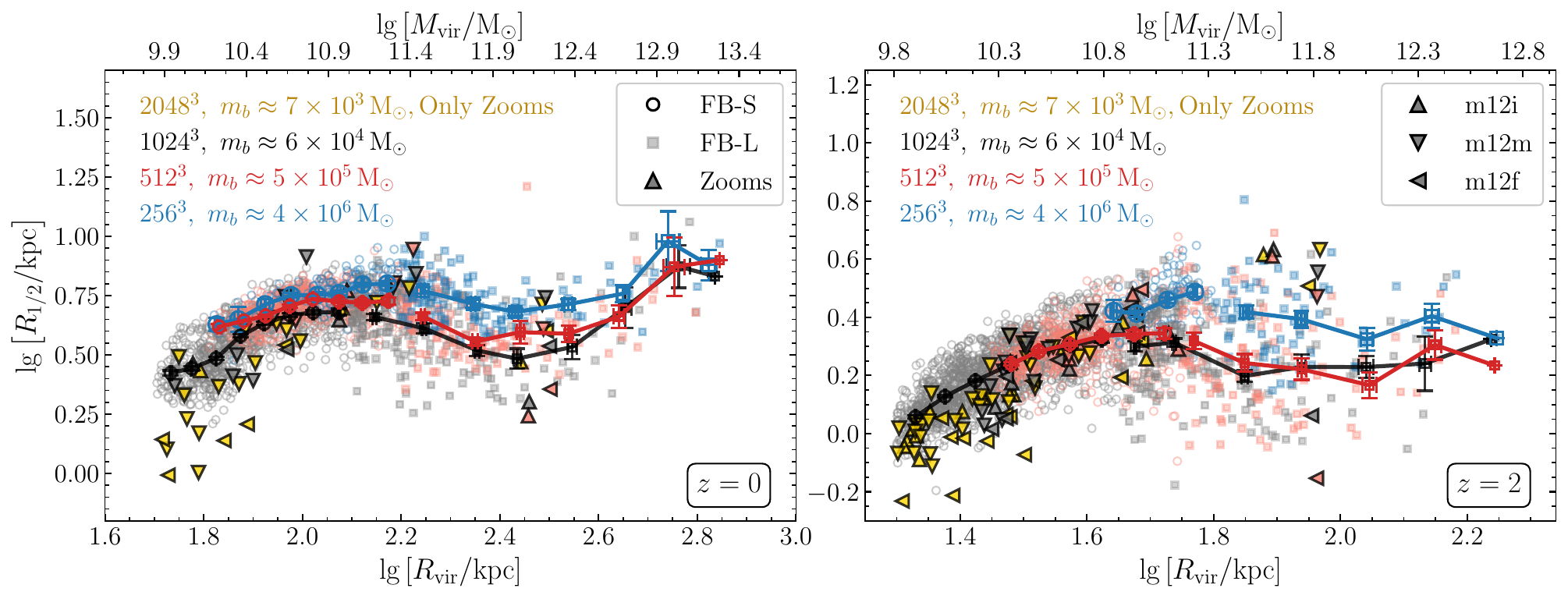}
    \caption{A convergence test of the GHSR at redshift $z=0$ (left) and $z=2$ (right), displaying the $256^3$ (blue), $512^3$ (red) and $1024^3$ (black) runs from FIREbox (FB). The open circles represent the galaxies within the selected $\rvir$, denoted `FB-S,' and the filled squares the galaxies with larger $\rvir$, denoted `FB-L.' Average sizes within each $\lg\rvir$ bin are overplotted with standard errors indicated by error bars. We also plot galaxy-halo pairs from FIRE-2 zoom simulations as triangles \citep{Wetzel2016,Hopkins2018}. These high resolution (gold triangles, $m_b\approx 7\times10^{3}\, \msun$) zoom-in simulations have an 8 times higher peak resolution than the FIREbox pathfinder ($1024^3$). Medium resolution zoom-in runs (grey triangles; $m_{\rm b} \approx 6\times 10^4\, \msun$) have comparable mass resolution to the $1024^3$ run. FIREbox galaxies in moderately massive haloes ($M_{\rm vir}\sim{}10^{11-13}$ $M_\odot$ at $z=0-2$) have approximately constant size with increasing halo size. In this study, we focus on the `FB-S' galaxies. The figure shows that the $\rhalf-\rvir$ relation appears converged at both $z=0$ and $z=2$ for the FIREbox pathfinder run.}
    \label{fig:convergence}
\end{figure*}

\begin{table}
 \caption{Selection and convergence criteria of the galaxies from the FIREbox 1024$^3$ simulation.}
 \label{tab:Selection}
 \begin{center}
 \begin{tabular}{cccccc}
  \hline
  $z$ & $\lg [\rvir/{\rm kpc}]$ & $\lg [M_{\rm vir}/\msun]$ & $N_{\rm FB-S}$ & $N_{\rm small}$ & $N_{\rm FB-L}$ \\
  (1) & (2) & (3) & (4) & (5) & (6) \\
  \hline
  0 & [1.70, 2.10] & [9.80, 11.00] & 826 & 47 & 189 \\
  1 & [1.50, 1.80] & [9.90, 11.25] & 1092 & 88 & 284 \\
  2 & [1.30, 1.65] & [9.80, 10.90] & 1373 & 49 & 210 \\ 
  3 & [1.20, 1.55] & [9.85, 10.95] & 1160 & 80 & 121 \\
  4 & [1.10, 1.45] & [9.85, 10.90] & 801 & 59 & 57 \\
  5 & [1.00, 1.35] & [9.80, 10.85] & 488 & 21 & 32 \\
  \hline
 \end{tabular}
 \end{center}
 \parbox{\columnwidth}{
  \footnotesize{
  (1) Redshift; (2) and (3) Ranges of the halo virial radius and mass respectively for the `FB-S' galaxies; (4) Number of galaxy/halo pairs included in the analysis; (5) and (6) Numbers of objects excluded because their halo sizes are below and above the virial radius ranges (column 2). 
  } 
 }
\end{table}

This paper only considers ``central galaxies", galaxies that form in main haloes (i.e., not satellite galaxies nor subhaloes). These galaxies are the most massive within $\rvir$ of the host halo and dominate the baryonic processes of the halo. Galaxy sizes and masses are calculated from the \texttt{disks} files provided by AHF. We consider the total galaxy radius to be $10\%$ of the halo virial radius \citep{Price2017}.At lower redshifts \citet{Hopkins2018,Samuel2020} find that satellites can exist within $0.1\rvir$ for massive galaxies, but this does not affect our sample of low mass galaxies. When including all stellar material within $0.2\rvir$, the same qualitative results hold (see the last paragraph of \S~\ref{sec:GHSR} for the results). Additionally, see Appendix~\ref{app:iter} for the results using an iterative $\rhalf$ calculation starting from all stars within $0.15\rvir$ \citep{Hopkins2018}.

We calculate the galaxy stellar mass $M_\star \equiv M_\star(<0.1\rvir)$ by linearly interpolating in $\log-\log$ space between radii $r$ and the cumulative stellar mass $M_\star(<r)$. Subsequently the inverse interpolation of $M_\star(<r)$ at $0.5M_\star$ yields the three-dimensional spherical half-stellar mass radius $\rhalf$ (see Appendix~\ref{app:iter} for the cumulative stellar mass radial profiles for the $1024^3$ `FB-S' galaxies at redshifts $z=0,\ 2$). The average stellar particle mass in the $1024^3$ simulation is $\sim 3\times 10^4\, \msun$ at $z=2$ (roughly half the baryonic mass resolution listed in \S~\ref{sec:NR} due to stellar mass loss). We consider the galaxies resolved when $N_\star(<0.1\rvir) \gtrsim 300$, implying $M_\star > 10^7\, \msun$ in the $1024^3$ simulation. We maintain this $N_\star$ criterion for all resolutions, meaning that the $M_\star$ lower limit in lower resolution simulations increases by factors of 8. 

Of the resolved galaxies, we study a selected range of halo sizes that depends on redshift and simulation resolution. At each redshift, we define a lower limit on $\rvir$ such that the number of haloes in each $\rvir$ bin decreases with increasing halo size. This lower limit on the halo size translates to a nearly constant lower limit on the halo mass of $M_{\rm vir} \gtrsim 10^{9.8}\, \msun$ since $z=5$. This lower limit cuts out a small percentage ($\sim4-8\%$, see Table~\ref{tab:Selection} column 5) of additional objects at each redshift. When we instead employ a looser criterion of a minimum halo mass of $M_{\rm vir} > 10^9\, \msun$, we only exclude a few additional haloes ($<5$ at each redshift), and the results remain qualitatively consistent. At large $\rvir$, the FIREbox galaxy sizes no longer increase with increasing halo size, which disagrees with observational results at these halo sizes \citep{Kravtsov2013,Huang2017,Zanisi2020}. At a fixed halo mass at $z=0$, these galaxies also have slightly larger stellar masses than is expected from the SHMR (see Feldmann et al., in prep for more details). Moreover, these galaxies have more rotational support ($v_{\rm rot} / v_{\rm circ} \sim 0.4$ at $z=0$; see \S~\ref{sec:galprop} for more details) than the galaxies at smaller $\rvir$ ($v_{\rm rot} / v_{\rm circ} \sim 0.1-0.2$). Lower-mass galaxies are more dispersion supported both for isolated dwarf galaxies in the Local Group \citep{Wheeler2017} and in the FIRE-1 zoom simulations \citep{Wheeler2015,Wheeler2017}. Hereafter, we denote the galaxies in the large-size regime as `FB-L', and the fiducial sample within the selected halo size range as `FB-S'. Due to the differences in the GHSR and amount of dispersion or rotational support between the `FB-L' and `FB-S' samples, we focus on the low-mass `FB-S' galaxies here. We plan to study the rotation supported, higher-mass `FB-L' galaxies in more detail in future work. Table~\ref{tab:Selection} summarises these limits and gives the number of galaxy/halo pairs at each redshift.

Figure~\ref{fig:convergence} displays the GHSR at redshifts $z=0$ and $z=2$ for the $256^3$, $512^3$ and $1024^3$ FIREbox simulations for galaxies residing in both small (`FB-S') and large haloes (`FB-L'). We also compare the predictions by FIREbox with recent FIRE-2 zoom simulations \citep{Wetzel2016,Hopkins2018}. The latter reach a mass resolution up to $\sim{}8\times$ higher than FIREbox pathfinder run allowing us to check for resolution effects. We also show results from 8 and 64 times lower resolution re-runs of these zoom-ins. The figure shows that the $\rhalf-\rvir$ relation appears well converged at both $z=0$ and $z=2$ for the $1024^3$ FIREbox run. However, \citet{Hopkins2018} shows that galaxy properties, including size, may still change when the mass resolution changes from $m_b\approx 6\times 10^{4}\, {\rm \msun}$ to $m_b\approx 7\times10^{3}\, {\rm \msun}$. Interestingly, the `FB-L' galaxies shows a prominent turn-over, i.e., a decrease in galaxy size with increasing halo size for $M_{\rm vir}\sim{}10^{11}-10^{12}\, \msun$ haloes. They also show an increased scatter. A few of the galaxies from the highest resolution zoom simulations have smaller sizes than the `FB-S' galaxies, which stems from the $M_\star$ criterion. At a fixed $\rvir$, the zoom simulations can resolve galaxies with $8\times$ lower stellar masses than the FIREbox pathfinder simulation. However, the `FB-S' sample contains many more galaxies larger than the zoom simulations, facilitating statistical analysis across a range of halo sizes.

In summary, the baryonic particles within an AHF defined dark matter halo must meet these criteria to be included in this analysis:
\begin{enumerate}
\item The host halo must be a main halo, i.e. not a proper subhalo.
\item The galaxy must be resolved with $N_\star \gtrsim 300$.
\item The halo's virial radius must be in the selected range outlined in Table~\ref{tab:Selection}.
\end{enumerate}

\section{Comparisons to Observations} \label{sec:comp}

\begin{figure}
    \includegraphics[width=\columnwidth]{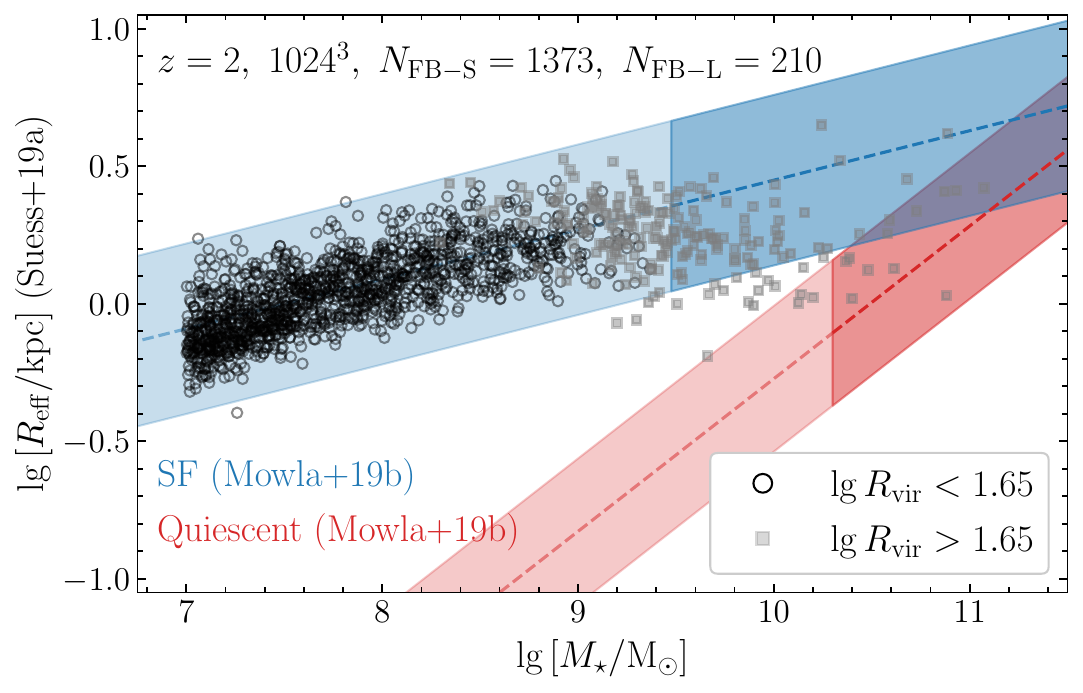}
    \caption{A comparison to the $R_{\rm eff}-M_\star$ relation obtained from CANDELS and COSMOS-DASH \citep{VanDerWel2014,Mowla2019}. Assuming that our 3D half-stellar mass radius $\rhalf$ has on average the same value in 2D projected space \citep{Ven2021}, then we convert our $\rhalf$ to a projected half-light radius $R_{\rm eff}$ \citep[][see text for more details]{Suess2019a}. Similarly to Figure~\ref{fig:convergence}, the open circles and filled squares represent the `FB-S' and `FB-L' galaxies at $z=2$, respectively. The dark shaded areas mark the mass range above the completeness limits of \citet{Mowla2019}, and the lighter regions show the relations extrapolated to lower masses. The completeness limit of \citet{Suess2019a} is $M_\star \gtrsim 10^{10}\, \msun$. The `FB-S' galaxies lies along the extrapolated relation for star forming galaxies.}
    \label{fig:Suess}
\end{figure}

\begin{figure}
    \includegraphics[width=\columnwidth]{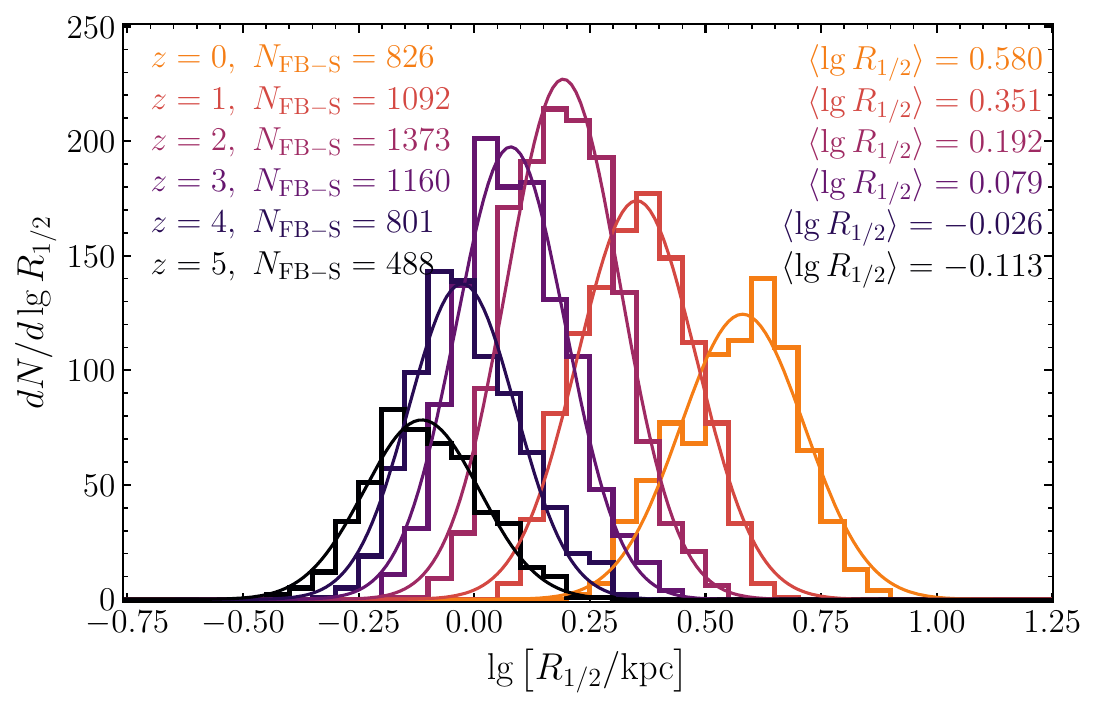}
    \caption{The galaxy size distribution for the `FB-S' galaxies ($M_\star \sim 10^{7-9}\, \msun$) at $z=0-5$. The distribution is well described by a log-normal at each redshift. The distribution parameters change with redshift, however. The Gaussian fitted means $\langle \lg[\rhalf/\kpc]\rangle$ are listed in the upper right corner. The average galaxy size increases with cosmic time (decreases with redshift and lookback time), qualitatively agreeing with observations \citep{Shibuya2015}. We find an approximately constant standard deviation $\sigma_{\lg \rhalf} = 0.12$ [dex] since $z=5$, also in qualitative agreement with \citet{Shibuya2015}.
    }
    \label{fig:zall_rhalf_lognormaldist}
\end{figure}

In Figure~\ref{fig:Suess}, we compare our galaxy size-stellar mass relation to observations at $z=2$ from the Cosmic Assembly Near-infrared Deep Extragalactic Legacy Survey (CANDELS) / 3D-HST \citep{VanDerWel2014} and Cosmic Evolution Survey (COSMOS) / Drift And SHift (DASH) \citep{Mowla2019} surveys. We assume that on average for a random projection, the intrinsic three-dimensional half-stellar mass radius $\rhalf$ is similar to the projected two-dimensional half-mass radius $R_{\rm mass}$ \citep[][see their text for how this approximation varies with intrinsic ellipsoidal axis ratios and for comparisons between $R_{1/2}$ and $R_{\rm eff}$]{Ven2021}. We convert $R_{\rm mass}$ to the 2D half-light radius $R_{\rm eff}$ using the empirical fit from \citet[][their Table 2 fit using $M_\star$ with a completeness limit of $M_\star > 10^{10}\, \msun$ at $1 < z < 2.5$]{Suess2019a}. Converting between $\rhalf$ and $R_{\rm eff}$ typically introduces a smaller scatter than the intrinsic scatter of the galaxy size-stellar mass relation \citep{Price2017,Genel2018,Ven2021}.

Extrapolating to lower masses where necessary, the `FB-S' galaxies lie along the extrapolated galaxy size-stellar mass relation for star-forming galaxies, while the `FB-L' galaxies turnover and lie partly between the relations for star-forming and quiescent galaxies. Recent observational \citep{Lange2016,Nedkova2021,Prole2021,Kawinwanichakij2021} and numerical \citep{Genel2018,Tremmel2020,Sales2020} studies at $z=0$ suggest that the relation for star-forming galaxies either continues or becomes flatter in the low-mass regime, while the quiescent relation flattens out for low-mass galaxies. There are no observational studies of dwarf galaxies at $z=2$. See Appendix~\ref{app:GSSMR} for more details regarding the galaxy size-stellar mass relation, and for the figures at $z=0$ (Figure~\ref{fig:GSSMR}). Briefly, at $z=0$ the FIREbox galaxies follow the galaxy size-stellar mass relation for star-forming galaxies from \citet{Nedkova2021}, but they are systematically larger by $\sim 0.3-0.5\, [{\rm dex}]$ at a fixed stellar mass. The value of the discrepancy depends on the correction factor employed from \citet{Suess2019b}. We emphasize that Figures~\ref{fig:Suess} and \ref{fig:GSSMR} should be understood as illustrations rather than proper comparisons with observations; we plan to analyse synthetic images of FIREbox galaxies using radiative transfer \citep[e.g.,][]{Liang2019,Liang2021} in future work.

\citet{Shibuya2015} find that a log-normal distribution well approximates the $R_{\rm eff}$ distribution in their combined legacy dataset of 3D-HST, CANDELS, Hubble Ultra Deep Fields (HUDF) 09+12 and the Hubble Frontier Fields (HFF) surveys. While their sample of star-forming and Lyman break galaxies focuses on more massive galaxies ($M_\star \sim 10^{8.5-11}\, \msun$ since $z=6$), our $\rhalf$ distribution qualitatively agrees. Figure~\ref{fig:zall_rhalf_lognormaldist} shows our best fitting Gaussians to the histograms of $\lg \rhalf$ in our final samples at each redshift. We create Q-Q plots for the fits to each redshift (omitted here), which indicate that a Gaussian well approximates each snapshot's $\lg \rhalf$ distribution. However, the Shapiro-Wilks tests and chi-squared $p$-values for normal distributions indicate that the distributions are likely not perfectly log-normal. Further, \citet{Shibuya2015} find that the average galaxy size decreases significantly towards higher redshift whilst maintaining a roughly constant standard deviation. Figure~\ref{fig:zall_rhalf_lognormaldist} demonstrates that `FB-S' galaxies display the same behaviour.

We find an approximately constant standard deviation $\sigma_{\lg \rhalf} = 0.12$ [dex] since $z=5$ in qualitative agreement with \citet{Shibuya2015}. However, their value for the scatter is somewhat higher at $\sigma \sim 0.2-0.3$ [dex]. Their sample of galaxies covers a broad range of luminosities, so it is not surprising that their measured scatter is slightly higher than ours. Additionally, \citet{VanDerWel2014} find a scatter $\sigma \sim 0.15-0.22$ [dex], in closer agreement to our measured value. We note that our galaxy sample changes across the different redshifts due to the minimum stellar mass criterion. Thus we are reporting the population average and not the progenitor average, which is more similar to observations.

\section{The Galaxy-Halo Size Relation} \label{sec:GHSR}

\begin{table*}
 \caption{GHSR at each and all redshifts from the FIREbox $1024^3$ `FB-S' galaxies}
 \label{tab:GHSR}
 \begin{center}
 \begin{tabular}{cccccc}
  \hline
  $z$ & $\alpha_{\rm GHSR}$ & $\beta_{\rm GHSR}$ & $\langle\sigma\rangle$ & $\alpha_{\sigma}$ & $\beta_\sigma$ \\
  (1) & (2) & (3) & (4) & (5) & (6) \\
  \hline
  $0$ & $0.929\pm0.033$ & $0.051\pm0.002$ & $0.089$ & $0.041\pm0.035$ & $0.029\pm0.309$ \\
  $1$ & $1.020\pm0.032$ & $0.052\pm0.001$ & $0.082$ & $0.097\pm0.037$ & $0.024\pm0.019$ \\
  $2$ & $0.939\pm0.024$ & $0.055\pm0.001$ & $0.080$ & $0.144\pm0.023$ & $0.021\pm0.005$\\
  $3$ & $0.909\pm0.027$ & $0.054\pm0.001$ & $0.082$ & $0.189\pm0.029$ & $0.019\pm0.004$ \\
  $4$ & $0.838\pm0.036$ & $0.051\pm0.002$ & $0.089$ & $0.262\pm0.033$ & $0.017\pm0.003$\\
  $5$ & $0.954\pm0.047$ & $0.055\pm0.004$ & $0.089$ & $0.189\pm0.039$ & $0.021\pm0.007$ \\
  \hline
  mean$^a$ & $0.934\pm0.054$ & $0.053\pm0.002$ & $0.084$ & $0.150\pm0.067$ & $0.022\pm0.004$ \\
  all$-z^b$ & $0.894\pm0.005$ & $0.053\pm0.001$ & $0.085$ & $0.015\pm0.005$ & $0.033\pm0.005$ \\
  \hline
 \end{tabular}
\end{center}
 \parbox{\textwidth}{
  \footnotesize{
  $^a$The averages weighted by the number of objects at each redshift. \\ 
  $^b$The combined sample of all redshift snapshots, treating objects from different snapshots equally. \\
  (1) Redshift; (2) and (3) Power-law index and normalisation for the GHSR from Equation~\eqref{eqn:pl}; (4) Average scatter in the GHSR; (5) and (6) Power-law index and normalisation for the GHSR scatter as a function of $\lg\rvir$ bin.
  }
 }
\end{table*}

\begin{figure}
    \includegraphics[width=\columnwidth]{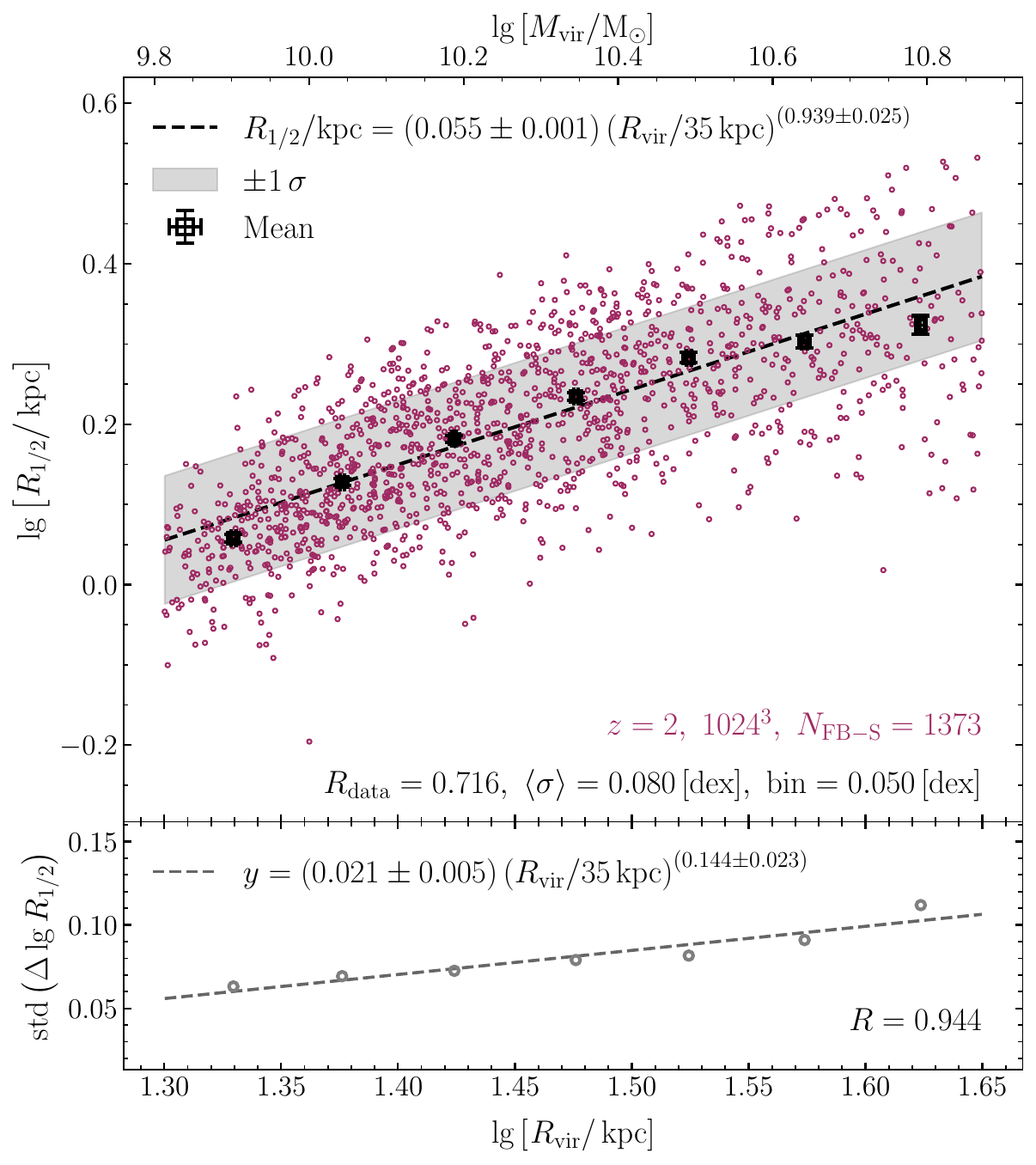}
    \caption{The GHSR relation for the galaxies at $z=2$ in FIREbox.
    \textit{Top Panel}: The open black squares represent the means $(\langle\lg \rvir\rangle, \langle\lg \rhalf\rangle)$ and standard errors within each $\lg \rvir$ bin of width 0.050 [dex]. The gray region represents $\pm 1\sigma$ scatter above and below the best fitting power-law (black dashed line). The least squares best fitting equation to the data is included in the upper left, and the Pearson correlation coefficient $R_{\rm data}$ and average scatter $\langle \sigma \rangle$ are shown in the bottom center. The slope of the best fitting power-law is approximately linear ($\alpha_{\rm GHSR} = 0.939\pm0.025$), agreeing with \citet{Kravtsov2013,Huang2017,Somerville2018}.
    \textit{Bottom Panel}: The standard deviation of the residuals for each $\lg \rvir$ bin from the GHSR in the upper panel. The least squares best fitting equation is included in the upper left, and its Pearson correlation coefficient is shown in the lower right, suggesting that the scatter slightly increases with $\rvir$.}
    \label{fig:z2_1024}
\end{figure}

\begin{figure}
    \includegraphics[width=\columnwidth]{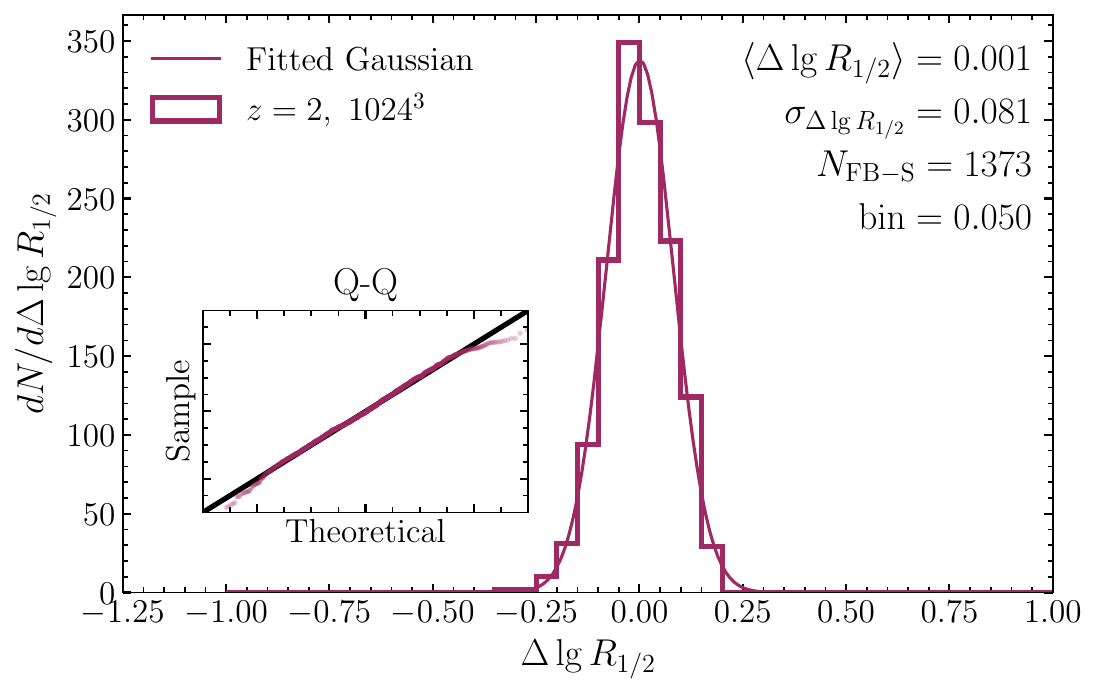}
    \caption{The Gaussian fit to the GHSR residuals $\Delta \lg \rhalf$ at $z=2$. The best fitting parameters are in the upper right corner, and the binwidth is 0.050 in $\lg$ space. The Q-Q plot in the lower left plot indicates that a Gaussian well approximates the $\Delta \lg \rvir$ distribution.}
    \label{fig:Residuals_lognormdist_z2}
\end{figure}

\begin{figure}
    \includegraphics[width=\columnwidth]{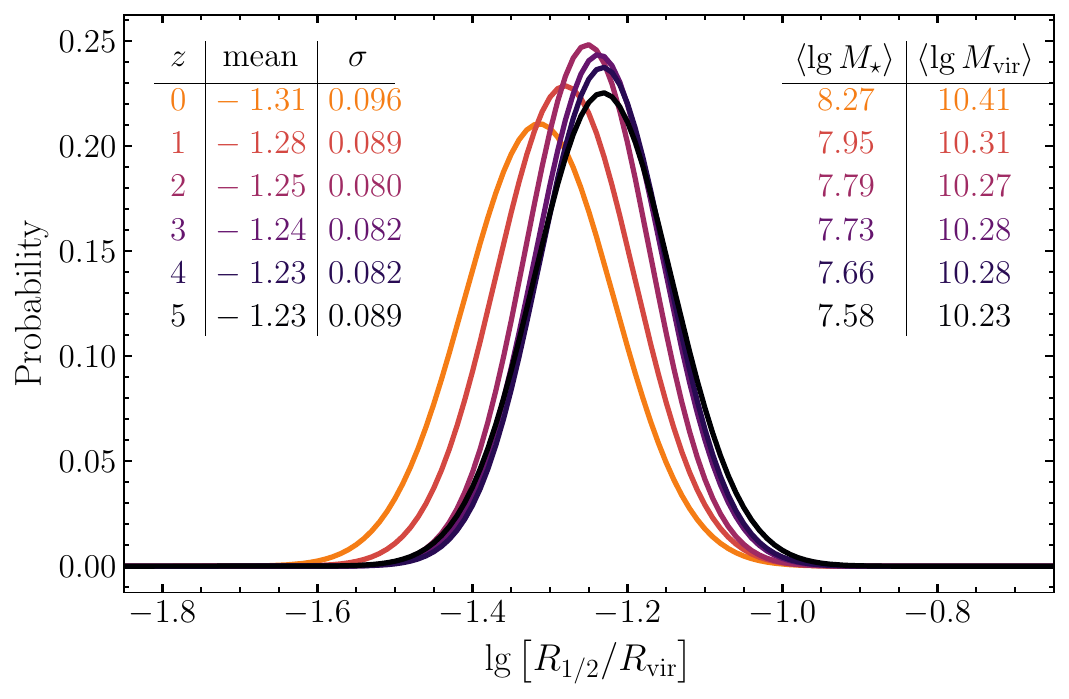}
    \caption{The log-normal probability distributions of the galaxy to halo size ratio of the FIREbox galaxies at different redshifts. We omit the histograms for clarity. The best fitting Gaussian parameters are shown in the upper left corner, and the average galaxy stellar and halo virial masses are included in the upper right corner. There is little redshift evolution in either the scatter ($\sigma \sim{}0.08-0.10)$ or the mean of this size ratio ($\rhalf / \rvir \sim 0.049-0.059$) since redshift $z=5$.}
    \label{fig:lg-rhalf-rvir_probdist_zall_1024}
\end{figure}

\begin{figure}
    \includegraphics[width=\columnwidth]{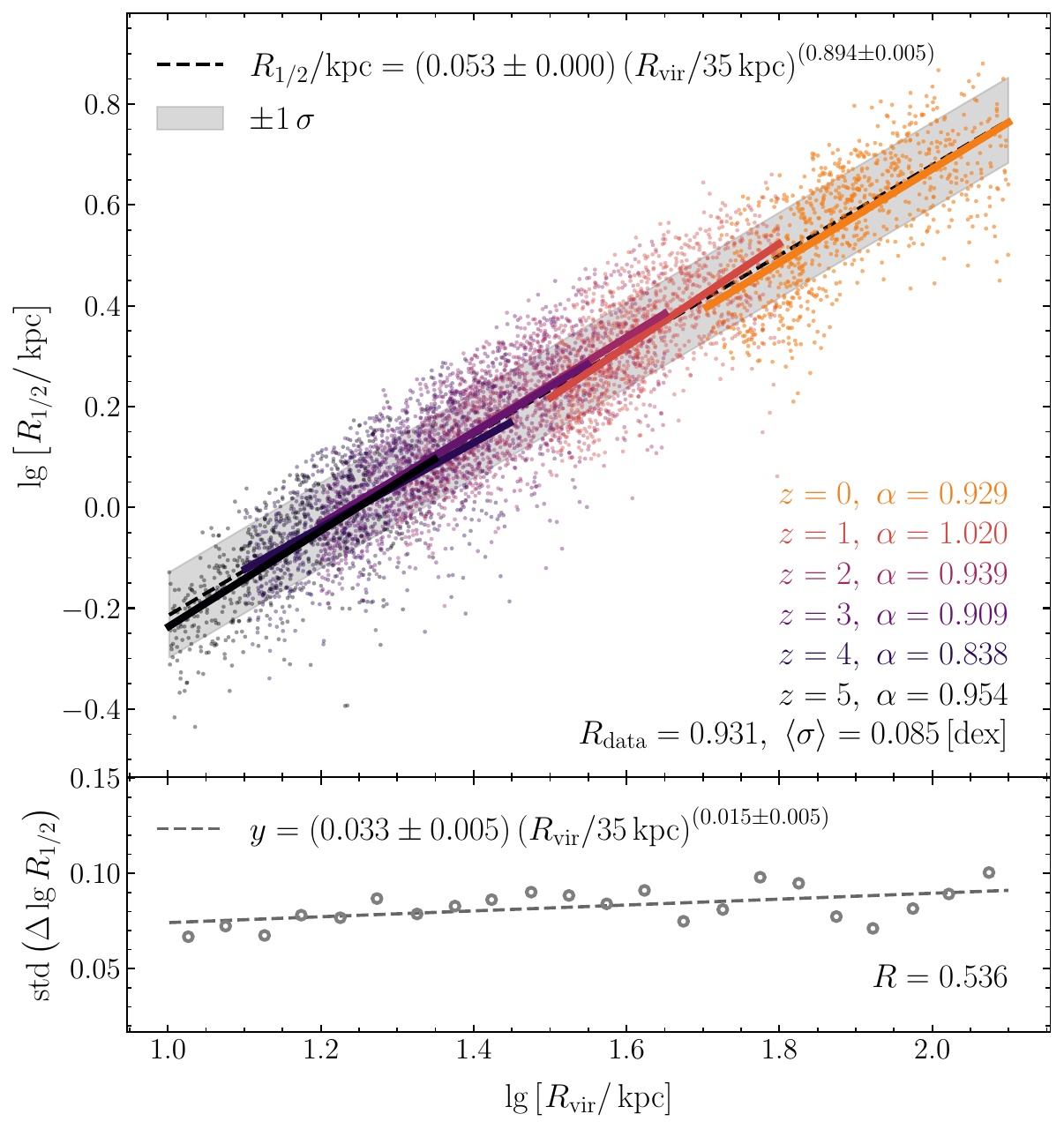}
    \caption{Similar to Figure~\ref{fig:z2_1024}, but we plot the best-fitting GHSR for each redshift as well as for the combined sample (all$-z$).
    \textit{Top Panel}: the color of the points represents their redshift (see legend), and the best-fitting power-law index $\alpha$ of the GHSR at that redshift is listed on the bottom right. The least squares fit to the combined (all$-z$) data is shown at the top of the panel. The gray region represents $\pm 1\sigma$ scatter above and below the best fitting power-law (black dashed line). The Pearson correlation coefficient $R_{\rm data}$ and the average scatter $\langle \sigma \rangle$ are listed at the bottom. The power-law relation of the combined data set is slightly sub-linear.
    However, the individual power-law relations are much closer to linear (except at $z=4$) suggesting that the size of a halo grows slightly faster than that of its central galaxy, qualitatively consistent with expectations from pseudo-evolution \citep{Diemer2013}.
    \textit{Bottom Panel}: The standard deviation of the residuals for each $\lg \rvir$ bin of width $0.050\, [{\rm dex}]$ from the GHSR in the upper panel. The equation of the best-fit and the Pearson correlation coefficient are shown at the top and bottom of the panel. The average scatter is small ($\langle \sigma \rangle <0.1$ [dex]) and only mildly dependent on halo size.}
    \label{fig:zallfits_1024}
\end{figure}

At each redshift, we construct the GHSR by fitting a power-law of the form 
\begin{equation} \label{eqn:pl}
    \rhalf\, [\kpc] = \beta\left(\dfrac{\rvir}{35\, \kpc}\right)^\alpha,
\end{equation}
where 35 kpc is the midpoint of the $\rvir$ range between $z=0-5$ in the `FB-S' haloes. The first and fiducial power-law is to the unbinned dataset. The $\alpha$ and $\beta$ values denote the power-law index and normalisation, respectively. Then we calculate the average scatter $\langle \sigma \rangle$ as the standard deviation of the residuals $\operatorname{std}{(\Delta \lg \rhalf)}$ from the fiducial power-law. Lastly, we fit a power-law to this scatter as function of $\lg \rvir$. 

Figure~\ref{fig:z2_1024} displays these results for $z=2$, where the top panel shows the GHSR and the bottom panel details the scatter (standard deviation of the residuals). We also fit power-laws to the binned means and medians and found similar best fitting functions. The best fitting power-law index of the GHSR at $z=2$, $\alpha_{\rm GHSR} = 0.939\pm0.025$, is close to unity. The approximately linear GHSR in FIREbox at Cosmic Noon in our target sample is qualitatively consistent with similar findings at $z=0$ \citep{Kravtsov2013} and at higher redshifts \citep{Huang2017,Zanisi2020}. However, we note that our sample consists of smaller and lower-mass galaxies than in those observational studies. The scatter in the GHSR of the `FB-S' galaxies exhibits a moderate increase with $\rvir$, as indicated by the non-vanishing power-law index $\alpha_\sigma = 0.144\pm0.023$. It is possible that the decreasing number of galaxies and known turnover effects at larger $\rvir$ cause the sub-linearities and increasing scatter.

We show the quality of our power-law fits by binning the residuals and fitting a Gaussian distribution to the resulting histogram, as Figure~\ref{fig:Residuals_lognormdist_z2} shows for $z=2$. The average value of the residuals $\langle \Delta \lg \rhalf \rangle$ is null, and the standard deviation $\sigma_{\Delta\lg\rhalf}$ is consistent with the average scatter in the GHSR. A Shapiro-Wilks test and the chi-squared $p$-value for a normal distribution indicate that the GHSR residuals $\Delta \lg \rhalf$ do not form a perfect normal distribution, but the Gaussian fit is a good approximation.

Table~\ref{tab:GHSR} summarises the main parameters for the GHSR and residuals for each of the analyzed redshifts. See Table~\ref{tab:GHSR_Hopkins} for the same table using the iterative galaxy definition \citep{Hopkins2018}. Given that the GHSR is consistent with being linear (column 2), the normalisation stays roughly constant (column 3), and the average scatter does not vary with redshift (column 4), we conclude that the GHSR in FIREbox is approximately constant since $z=5$. We calculate the weighted averages of each of the redshifts considered and find that the power-law index $\alpha_{\rm GHSR} = 0.934\pm0.054$ is approximately linear and the scatter $\langle\sigma\rangle = 0.084$ is constant with individual snapshots.

Figure~\ref{fig:lg-rhalf-rvir_probdist_zall_1024} displays the probability distributions of the galaxy to halo size {\it ratio} at each redshift. We fit log-normal distributions to histograms of $\lg[\rhalf / \rvir]$, but we do not show the histograms for clarity. There is little redshift evolution in either the mean or the scatter of this size ratio, and we find typical values of $\rhalf / \rvir \sim 0.05$, which is higher than \citet{Kravtsov2013,Shibuya2015,Somerville2018} who find values closer to $\sim 0.02$. However, these studies focus on more massive galaxies than the low-mass centrals analyzed here. As Figure~\ref{fig:convergence} shows, low- and high-mass FIREbox galaxies do not necessarily follow the same GHSR. \citet{Somerville2018} find a weak redshift dependence on this ratio (see their Figure 12). However, this dependence decreases with decreasing halo mass, and our sample is more than an order of magnitude less massive than their least massive bin.

We also combine all galaxy-halo pairs from each redshift and construct a GHSR from this total dataset, treating objects from different redshifts equally. This means that the definition of $\rvir$ is a function of redshift, because $\rvir$ depends on the top hat collapse factor times the background density $\Delta\rho_{\rm back} = \Delta(z)\rho_{\rm back}(z)$ \citep{Bryan1998}. Figure~\ref{fig:zallfits_1024} displays these results, where the top panel shows the GHSR and the bottom panel the standard deviation of the residuals. The combined sample spans over an order of magnitude in halo size, and there is a ``discretely smooth" transition between the six redshift snapshots. The best fitting power-law index decreases to $\alpha = 0.894\pm0.005$, indicating that this combined sample is slightly sub-linear. This contrasts with the individual and averaged GHSRs, whose power-law index remains consistent with linear at $\alpha = 0.934\pm0.054$. At any given instant in time the GHSR is linear, but the size of a given halo typically grows slightly faster than the size of its central galaxy over much of cosmic history. This sub-linearity and the slight decrease of $\langle \lg[\rhalf / \rvir]$ with cosmic time (Figure~\ref{fig:lg-rhalf-rvir_probdist_zall_1024}) are qualitatively consistent with expectations of pseudo-evolution of halo sizes \citep{Diemer2013}. However, these low-mass central galaxies remain star-forming until $z=0$, so their galaxy sizes continue growing as well. Nonetheless, it appears that halo sizes grow slightly quicker than galaxy sizes since $z=5$. Hence, the average $\langle \lg[\rhalf / \rvir]\rangle$ in Figure~\ref{fig:lg-rhalf-rvir_probdist_zall_1024} shift leftward with cosmic time (decreasing $z$), and the power-law fit to the all-z sample is less linear than the fits to each individual redshift in Figure~\ref{fig:zallfits_1024}.

As Figure~\ref{fig:zallfits_1024} shows and Table~\ref{tab:GHSR} summarises, the GHSR at each redshift, as well as the combined GHSR, have nearly constant scatter at $\langle \sigma \rangle \approx 0.08-0.09\, [{\rm dex}]$, suggesting that redshift is {\it not} a significant factor in explaining the scatter in the GHSR. We emphasize this point by explicitly accounting for a redshift dependence by fitting
\begin{equation} \label{eqn:threeparam}
    \lg \left[\rhalf / \kpc\right] = \alpha_{\rvir}\lg \left[\rvir / \kpc\right] + \alpha_{1+z}\lg\left[1+z\right] + \beta. 
\end{equation}
The power-law index for redshift $\alpha_{1+z} = 0.049 \pm 0.013$ is approximately 0, and the index for virial radius increases from $\alpha_{\rm GHSR} = 0.894\pm0.005$ (Equation~\ref{eqn:pl}; Figure~\ref{fig:zallfits_1024}) to $\alpha_{\rvir} = 0.939\pm0.013$. The average scatter $\langle \sigma \rangle = 0.085$ remains exactly the same as that of Figure~\ref{fig:zallfits_1024}. Moreover, the log-Gaussian fits to the residuals have the same standard deviation of $\sigma_{\Delta \lg R_{1/2}} = 0.086\, [{\rm dex}]$. Hence, redshift is not a significant factor in explaining the scatter in the GHSR in these low mass galaxies.

We repeat this analysis for $R_{25}$ and $R_{80}$, i.e. the radii containing $25\%$ and $80\%$ of the stellar mass within $0.1\rvir$ respectively \citep{Miller2019,Mowla2019a}. The scatter changes from $\langle\sigma\rangle \approx 0.08-0.09\, [{\rm dex}]$ with $\rhalf$ to $\approx 0.05$ and $\approx 0.11$ [dex] with $R_{25}$ and $R_{80}$ respectively. 
\textit{We also repeat this analysis for $\rhalf$ using all stars within $0.2\rvir$, and we find that the average scatter of the GHSR increases to $\langle\sigma\rangle \approx 0.12-0.13\, [{\rm dex}]$. When including redshift in the two-parameter fit, there is a larger redshift power-law index $\alpha_{1+z}$ to $\rhalf$ from equation~\eqref{eqn:threeparam}: $\alpha_{\rvir} = 0.885\pm0.01,\ \alpha_{1+z} = 0.128\pm0.018$. However the overall scatter remains constant.}
Lastly, we repeat this analysis for the iterative definition of galaxy size $\rhalf$ described in \citet[][see Appendix \ref{app:iter} for more details]{Hopkins2018}. Again, the scatter increases to $\langle\sigma\rangle = 0.146\, [{\rm dex}]$, and there is a larger redshift power-law index $\alpha_{1+z}$ in the two-parameter fit: $\alpha_{\rvir} = 0.925\pm0.022,\ \alpha_{1+z} = 0.175\pm0.023$. The scatter still does not decrease by including redshift in the two-parameter fit. For each definition of the galaxy size, the GHSRs are approximately linear, and the scatter remains unchanged when including redshift as a second parameter for galaxy size.

\section{What sets the scatter in the GHSR?} \label{sec:scat}

\begin{figure*}
    \includegraphics[width=\textwidth]{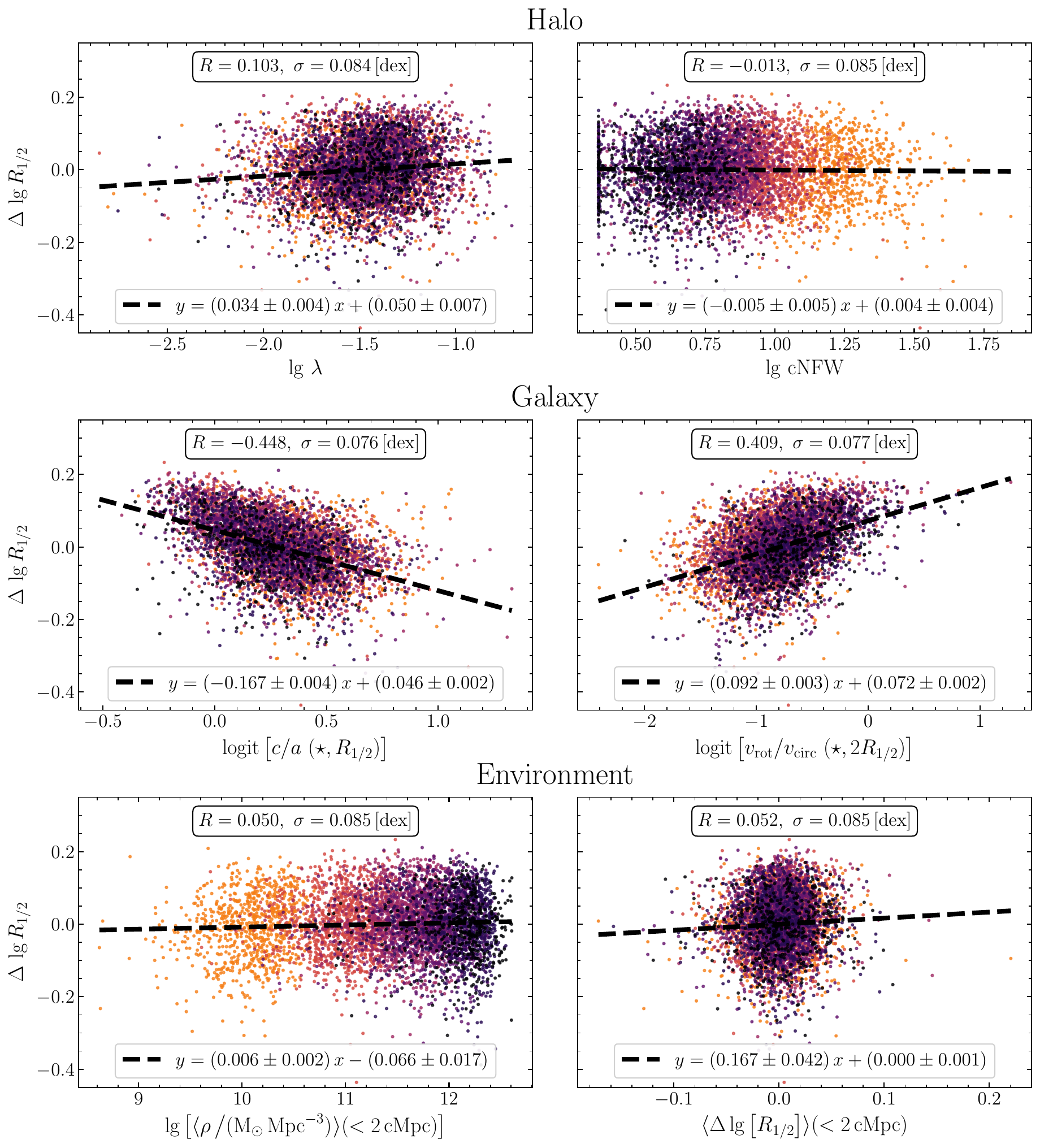}
    \caption{Six example figures graphing the residuals $\Delta \lg \rhalf$ (i.e., the scatter) in the galaxy-halo size relation (GHSR) versus two halo (top panels), galaxy (middle), and environment properties (bottom). The best fitting power-laws are shown with black dashed lines, and their respective equations, Pearson correlation coefficients and new scatter are given in each panel. These values are summarised for all properties in Table~\ref{tab:Param}. The average scatter of the GHSR is $\sigma_{\Delta \lg R_{1/2}} = 0.085$ [dex].
    \textit{Top Left}: The halo spin. While the best fiting slope is $8.5\sigma$ from 0, the scatter does not decrease.
    \textit{Top Right}: The halo concentration does not correlate with the GHSR scatter.
    \textit{Middle Left}: The shape of the stars correlates with GHSR scatter and deceases the overall scatter. At a fixed halo size, galaxies with stars that are less spherical (lower $c/a$) have larger sizes. This property significant in setting the GHSR scatter.
    \textit{Middle Right}: The ratio of stellar rotational velocity to the circular velocity at the half-mass radius. At a fixed halo size, galaxies with more rotational support (higher $v_{\rm rot} / v_{\rm circ}$) have larger sizes. This property is anti-correlated with the stellar shape $c/a$. 
    \textit{Bottom Left}: The average total density of all central neighbors within 2~cMpc does not correlate with the GHSR scatter. 
    \textit{Bottom Right}: The average GHSR residuals of all central neighbors (size conformity) within 2~cMpc. The average size conformity correlates weakly with the residuals, but the GHSR scatter does not decrease.}  
    \label{fig:paramplot6panel}
\end{figure*}

\begin{table*}
 \caption{Correlations of the `FB-S' GHSR's scatter with halo (top), galaxy (middle) and environment (bottom) properties at all considered redshifts.}
 \label{tab:Param}
 \begin{center}
 \begin{tabular}{lcccccc}
  \hline
  Parameter & $m$ & $\sigma_{m,0}$ & $b$ & $R$ & $\langle\sigma\rangle$  &  $\%\Delta\sigma$ \\
  (1) & (2) & (3) & (4) & (5) & (6) & (7) \\
  \hline
${\rm logit}\left[ b/a\ ({\rm halo}, \rvir) \right]$ & $0.013\pm0.003$ & $4.14$ & $-0.012\pm0.003$ & $0.055$ & $0.085\pm0.001$ & $0.1$ \\
${\rm logit}\left[ c/a\ ({\rm halo}, \rvir) \right]$ & $0.022\pm0.005$ & $4.43$ & $-0.012\pm0.003$ & $0.058$ & $0.085\pm0.001$ & $0.2$ \\
${\rm logit}\left[ c/b\ ({\rm halo}, \rvir) \right]$ & $0.004\pm0.004$ & $0.99$ & $-0.003\pm0.004$ & $0.013$ & $0.085\pm0.001$ & $0.0$ \\
${\rm logit}\left[ E \equiv \sqrt{1 - (b/a)^2}\ ({\rm halo}, \rvir) \right]$ & $-0.018\pm0.004$ & $4.37$ & $-0.001\pm0.001$ & $-0.058$ & $0.085\pm0.001$ & $0.2$ \\
${\rm logit}\left[ F \equiv \sqrt{1-(c/b)^2}\ ({\rm halo}, \rvir) \right]$ & $-0.005\pm0.005$ & $0.97$ & $-0.000\pm0.001$ & $-0.013$ & $0.085\pm0.001$ & $0.0$ \\
${\rm logit}\left[ T \equiv \left.\left(1 - \left(b/a\right)^2\right)\right/\left(1- \left(c/a\right)^2\right)\ ({\rm halo}, \rvir) \right]$ & $-0.007\pm0.002$ & $2.78$ & $0.001\pm0.001$ & $-0.037$ & $0.085\pm0.001$ & $0.1$ \\
$\lg\, \lambda$ & $0.034\pm0.004$ & $7.82$ & $0.050\pm0.007$ & $0.103$ & $0.084\pm0.001$ & $0.5$ \\
$\lg\, \lambda_e$ & $0.030\pm0.005$ & $6.58$ & $0.046\pm0.007$ & $0.086$ & $0.084\pm0.001$ & $0.4$ \\
$\lg\, {\rm cNFW}$ & $-0.005\pm0.005$ & $0.99$ & $0.004\pm0.004$ & $-0.013$ & $0.085\pm0.001$ & $0.0$ \\
$\lg\, \left[\sigma_v/\rm km\, s^{-1}\right]$ & $0.021\pm0.009$ & $2.28$ & $-0.038\pm0.017$ & $0.030$ & $0.085\pm0.001$ & $0.0$ \\
$\lg\, \left[\Delta\, {\rm COM}/{\rm kpc}\right]$ & $0.064\pm0.003$ & $19.32$ & $-0.039\pm0.002$ & $0.247$ & $0.082\pm0.001$ & $3.1$ \\
\hline
\boldmath${\rm logit}\left[ b/a\ (\star, R_{1/2}) \right]$ & \boldmath$-0.101\pm0.003$ & \boldmath$31.84$ & \boldmath$0.063\pm0.002$ & \boldmath$-0.388$ & \boldmath$0.078\pm0.001$ & \boldmath$7.8$ \\
\boldmath${\rm logit}\left[ c/a\ (\star, R_{1/2}) \right]$ & \boldmath$-0.167\pm0.004$ & \boldmath$38.00$ & \boldmath$0.046\pm0.002$ & \boldmath$-0.448$ & \boldmath$0.076\pm0.001$ & \boldmath$10.6$ \\
${\rm logit}\left[ c/b\ (\star, R_{1/2}) \right]$ & $-0.056\pm0.004$ & $14.54$ & $0.040\pm0.003$ & $-0.188$ & $0.083\pm0.001$ & $1.8$ \\
\boldmath${\rm logit}\left[ E \equiv \sqrt{1 - (b/a)^2}\ (\star, R_{1/2}) \right]$ & \boldmath$0.120\pm0.004$ & \boldmath$33.22$ & \boldmath$-0.021\pm0.001$ & \boldmath$0.402$ & \boldmath$0.078\pm0.001$ & \boldmath$8.4$ \\
${\rm logit}\left[ F \equiv \sqrt{1- (c/b)^2}\ ({\star}, R_{1/2}) \right]$ & $0.069\pm0.005$ & $15.06$ & $-0.006\pm0.001$ & $0.195$ & $0.083\pm0.001$ & $1.9$ \\
${\rm logit}\left[ T \equiv \left.\left(1 - \left(b/a\right)^2\right)\right/\left(1- \left(c/a\right)^2\right)\ ({\star}, R_{1/2}) \right]$ & $0.046\pm0.003$ & $18.40$ & $-0.013\pm0.001$ & $0.236$ & $0.082\pm0.001$ & $2.8$ \\
$\lg\left[\left. M_\star \right/ {\rm M}_\odot \right]$ & $-0.007\pm0.002$ & $4.01$ & $0.057\pm0.014$ & $-0.053$ & $0.085\pm0.001$ & $0.1$ \\
\boldmath${\rm logit}\left[ \left. M_\star(<R_{1/2})\right/M_{\rm bar}(<R_{1/2}) \right]$ & \boldmath$-0.047\pm0.002$ & \boldmath$26.27$ & \boldmath$-0.019\pm0.001$ & \boldmath$-0.336$ & \boldmath$0.079\pm0.001$ & \boldmath$5.8$ \\
${\rm logit}\left[ \left. M_\star(<R_{1/2})\right/M_{\rm tot}(<R_{1/2}) \right]$ & $-0.062\pm0.003$ & $22.86$ & $-0.074\pm0.003$ & $-0.289$ & $0.081\pm0.001$ & $4.3$ \\
$\lg\left[\left.M_{\rm gas}(<R_{1/2})\right/{\rm M}_\odot\right]$ & $0.029\pm0.002$ & $18.94$ & $-0.226\pm0.012$ & $0.249$ & $0.081\pm0.001$ & $3.1$ \\
${\rm logit}\left[ \left. M_{\rm gas}(<R_{1/2})\right/M_{\rm tot}(<R_{1/2}) \right]$ & $0.048\pm0.002$ & $20.29$ & $0.034\pm0.002$ & $0.267$ & $0.081\pm0.001$ & $3.6$ \\
$\lg\left[\left. M_{\rm bar}(<R_{1/2}) \right/ {\rm M}_\odot\right]$ & $0.027\pm0.002$ & $14.13$ & $-0.220\pm0.016$ & $0.183$ & $0.083\pm0.001$ & $1.7$ \\
${\rm logit}\left[ \left. M_{\rm bar}(<R_{1/2})\right/ M_{\rm tot}(<R_{1/2}) \right]$ & $0.025\pm0.003$ & $8.74$ & $0.011\pm0.002$ & $0.115$ & $0.084\pm0.001$ & $0.7$ \\
$\lg\left[\left. M_{\rm dm}(<R_{1/2})\right/{\rm M}_\odot\right]$ & $0.057\pm0.004$ & $15.80$ & $-0.491\pm0.031$ & $0.204$ & $0.083\pm0.001$ & $2.1$ \\
$\lg\left[\left. M_{\rm tot}(<R_{1/2})\right/{\rm M}_\odot\right]$ & $0.043\pm0.003$ & $13.80$ & $-0.376\pm0.027$ & $0.179$ & $0.083\pm0.001$ & $1.6$ \\
\boldmath${\rm logit} \left[v_{\rm rot} / v_{\rm circ}\ (\star, R_{1/2}) \right]$ & \boldmath$0.092\pm0.003$ & \boldmath$33.89$ & \boldmath$0.072\pm0.002$ & \boldmath$0.409$ & \boldmath$0.077\pm0.001$ & \boldmath$8.7$ \\
\hline
$\lg\left[\left.{\rm min}({\rm d})\right/{\rm kpc}\right]$ & $-0.016\pm0.003$ & $5.10$ & $0.036\pm0.007$ & $-0.067$ & $0.085\pm0.001$ & $0.2$ \\
$\lg\left[\left.{\rm min}(R_{\rm Hill})\right/{\rm kpc}\right]$ & $-0.013\pm0.003$ & $4.26$ & $0.027\pm0.006$ & $-0.056$ & $0.085\pm0.001$ & $0.2$ \\
$\lg\left[\langle n_{\rm gal}\left/{\rm Mpc}^{-3}\right.\rangle(<2\, {\rm cMpc})\right]$ & $0.003\pm0.002$ & $1.76$ & $-0.002\pm0.002$ & $0.024$ & $0.084\pm0.001$ & $0.0$ \\
$\lg\left[\langle\rho \left/({\rm M_\odot\, Mpc}^{-3})\right.\rangle(<2\, {\rm cMpc})\right]$ & $0.006\pm0.002$ & $3.80$ & $-0.066\pm0.017$ & $0.050$ & $0.085\pm0.001$ & $0.1$ \\
$\Delta \lg \left[R_{\rm 1/2,min(R_{\rm Hill})}\right]$ & $0.025\pm0.012$ & $2.13$ & $0.000\pm0.001$ & $0.028$ & $0.085\pm0.001$ & $0.0$ \\
$\langle\Delta \lg \left[R_{1/2}\right]\rangle(<2\, {\rm cMpc})$ & $0.167\pm0.042$ & $3.97$ & $0.000\pm0.001$ & $0.052$ & $0.085\pm0.001$ & $0.1$ \\
\hline
 \end{tabular}
 \end{center}
 \parbox{\textwidth}{
  \footnotesize{
  (1) Parameter used as the horizontal axis -- $\logit x \equiv \lg x / (1-x)$ for $x \in (0, 1)$; (2) Slope of the fit; (3) $\sigma$ from the slope being $0$ using the statistical error of $m$; (4) Vertical offset in the fit; (5) Pearson correlation coefficient; (6) Scatter in regression in the residuals versus parameter; (7) Percentage difference in scatter between the GHSR and the residual-parameter relation. 
  From top to bottom the sections are the halo, galaxy, and environment properties. \S~\ref{sec:scat} describes what each parameter is and how it is calculated. 
  Parameters that explain at least five percent of the scatter (column (7)) --  $\%\Delta\sigma > 5$ -- are in bold.
  }
 }
\end{table*}

The scatter in the GHSR is approximately constant since $z = 5$ in the low-mass central galaxies in FIREbox (`FB-S'). Because including redshift as another parameter to the GHSR does not decrease this scatter, we now investigate halo (\S~\ref{sec:haloprop}), galaxy (\S~\ref{sec:galprop}), and environment (\S~\ref{sec:envirprop}) properties that could determine this scatter. Specifically, we correlate the GHSR residuals with these physical properties. If there is a strong correlation \textit{and} the average scatter decreases, then this property is important in setting the GHSR scatter.

Figure~\ref{fig:paramplot6panel} exemplifies six of these scatter plots for the `FB-S' galaxies from all redshifts. The vertical axis shows the residuals -- i.e., the difference in $\log$ space of the true galaxy size $\rhalf$ and that predicted from the GHSR and the halo's $\rvir$ -- as a function of various properties. We employ a linear regression between the log residuals and functions of the properties. For physical properties -- e.g., halo spin $\lambda$, stellar mass $M_\star$, and local density $\langle \rho \rangle$ -- we fit the residuals $\Delta \lg \rhalf$ with the $\lg(x)$ of the physical property $x$; for fractional properties -- e.g., shape parameters $c/a$ and mass fractions $M_\star / M_{\rm bar}$-- we use the $\logit(x) \equiv \lg(x / (1-x))$ of the property. These functional choices ($\lg(x)$ for physical properties $x \in (0, \infty)$; $\logit(x)$ for fractional properties $x \in (0, 1)$) map their respective properties to all real numbers $(-\infty, \infty)$. For each of these residual-parameter figures, we study the statistical significance $\sigma_{m,0}$ of the slope $m$ from 0 and the percent reduction in scatter $\%\Delta\sigma$. We also calculate the vertical offset $b$ of the fit and the Pearson correlation coefficient $R$. Table~\ref{tab:Param} details these results for the `FB-S' galaxies for the halo (top), galaxy (middle), and environment (bottom) properties respectively. We also repeat this analysis at each redshift and find that all results are consistent with the combined `all-$z$' sample.

\subsection{Halo Properties} \label{sec:haloprop}

The top set of properties correlates the GHSR residuals with the halo properties, employing the AHF values of the properties at the virial radius $\rvir$ using all -- dark matter and baryonic -- interior particles. 

The first six halo properties describe the shape using the principal axes of the moment of inertia tensor, such that $a > b > c$. Namely, $b/a,\ c/a,\ {\rm and}\ c/b$ are the ratios of these axes, and the elongation $E$, flattening $F$, and triaxiality $T$ are derived from these ratios. In general as a halo becomes less spherical/more triaxial, the axis ratios decrease while the elongation, flattening, and triaxiality increase. Each of these six values is between 0 and 1, so we correlate the $\logit$ of each parameter with the GHSR residuals. While $\logit b/a$, $\logit c/a$, and $\logit E$ have slopes $>4\sigma$ from null, they and the other four properties do not decrease the scatter. Consequently, we conclude that the halo shape is not significant in setting the scatter in the GHSR.

Many theoretical and empirical works suggest that the halo spin and concentration are significant in setting the galaxy sizes, especially in rotationally supported disk galaxies \citep{Fall1980,Mo1998,Somerville2008,Diemer2015,Desmond2015,Desmond2017,Somerville2018}. We study the halo spin $\lg \lambda, \lg\lambda_e$ (top left panel of Figure~\ref{fig:paramplot6panel}) employing the definition of \citet{Bullock2001}:
\begin{equation} \label{eqn:spin}
\lambda = \frac{J}{\sqrt{2}M_{\rm vir}V_{\rm circ}\rvir},
\end{equation}
and the classical definition of \citet{Peebles1969}:
\begin{equation} \label{eqn:spin_e}
\lambda_e(\rvir) = \frac{J |E|^{1/2}}{GM_{\rm vir}^{5/2}},
\end{equation}
where $J = J(<r)$ is the angular momentum, $V_{\rm circ}$ is the circular velocity of all particles within the halo, $E$ is the total energy of the system and $G$ is Newton's gravitational constant. Again, while their slopes are $m\sim7\sigma$ from null, they do not decrease the scatter. 

Assuming a Navarro-Frank-White (NFW) dark matter profile \citep{Navarro1997}, we calculate the halo concentration using the definition from \citet[][see their \S~3, Eqs.~9-10 for more details]{Prada2012}. This concentration is also a proxy for the formation history of the halo \citep{Wechsler2002}. The fit to $\lg {\rm cNFW}$ has no significant correlation (top right panel of Figure~\ref{fig:paramplot6panel}). Because the concentration is linked to the halo formation history, this lack of correlation suggests that the halo formation time is not critical in the GHSR scatter.

We fit the centre-of-mass offset $\lg \Delta {\rm COM}$ -- defined as the distance between the halo's centre of mass and the halo's centre calculated as the densest cell. A larger offset implies some spherical asymmetry in the halo, such as massive satellites. Interestingly this property has the strongest correlation with the residuals with a slope that is $19\sigma$ from null, but the scatter only decreases by $\approx 3\%$. 

\textit{Thus, we conclude that none of the studied halo properties -- including the halo spin and the concentration -- significantly explain the scatter in the GHSR in our FIREbox sample of low-mass central galaxies over the past 12.5 billion years.}

In Appendix~\ref{app:dmo}, we also correlate the GHSR scatter with the difference between a property's value and the average value of all similar-massed haloes, the value of the cross-matched halo in the dark matter only (DMO) simulation, and the difference between the cross-matched halo value and the average value of all similar-massed DMO haloes. There are no significant reductions in the GHSR scatter for any of the studied properties. Thus all null-correlations between the halo properties and GHSR scatter are consistent, and we conclude that the halo properties are not important in setting the GHSR scatter in the `FB-S' galaxies.

\subsubsection{Halo Concentration and Spin} \label{sec:halocon_spin}

\begin{figure*}
    \includegraphics[width=\textwidth]{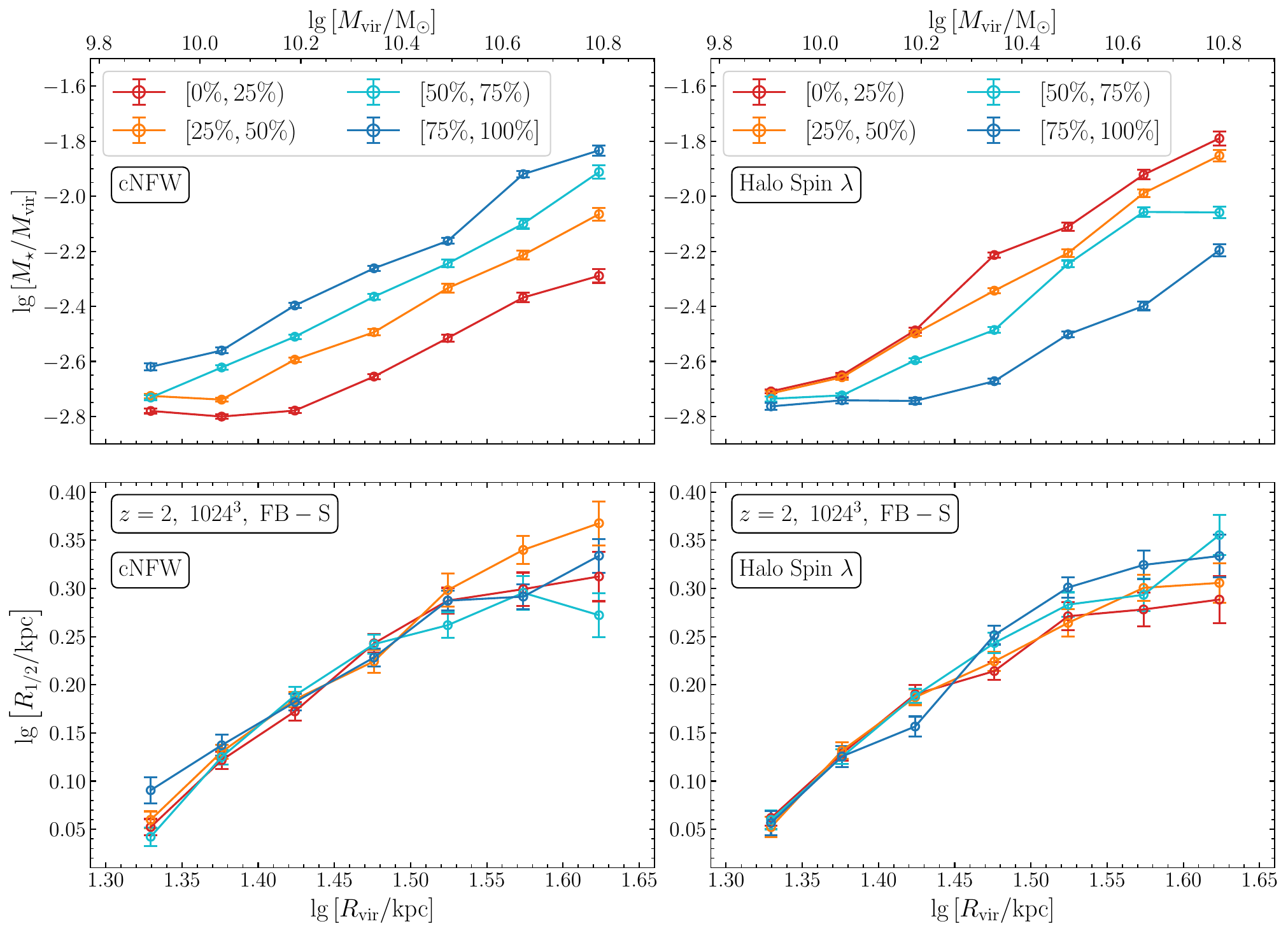}
    \caption{The stellar-halo mass (top row) and galaxy-halo size (bottom row) relations for the FIREbox galaxies at $z=2$, separated into equal sized quartiles by halo concentration cNFW (left column) and spin $\lambda$ (right panel). The $M_\star-M_{\rm vir}$ relation depends on the concentration and halo spin, while the $\rhalf-\rvir$ relation does not. However the largest, most massive haloes in in the GHSR may have a slight dependence on halo spin (bottom right panel). The same qualitative results hold at other redshifts. Thus, we suggest that for the low-mass, central, dispersion supported systems, the galaxy-halo size relation is just as or more fundamental than the stellar-halo mass relation.}
    \label{fig:Feldmann}
\end{figure*}

In Figure~\ref{fig:Feldmann}, we construct the stellar-halo mass (top row) and galaxy-halo size (bottom row) relations for the `FB-S' galaxies at $z=2$ (the same results hold at the other redshifts), where the sample is divided into four equal quartiles by halo concentration (left column) and spin (right column).  For the SHMR, at a fixed halo mass, haloes with a higher concentration have higher stellar masses (top left panel). The four SHMRs follow the same shape, but the high concentration curves have larger normalisations, appearing as vertical shifts. These results for the halo concentration agree with FIRE-1 results from \citet{Feldmann2019}. These vertical translations in the SHMR with concentration are equivalent to varying normalisations of the stellar density profiles \citep{Lilly2016}. The trend reverses for the halo spin; at a fixed halo mass, haloes with smaller spin have larger stellar masses (top right panel). For the `FB-S' galaxies, the concentration and the spin are significant properties in setting the SHMR.

However, there is no such distinction in the GHSR relations when separated into quartiles by either concentration or spin. The null result with concentration agrees with the semi-empirical model from \citet{Zanisi2021a} and disagrees with results from zoom-ins from \citet{Jiang2019}, although both studies focused on more massive galaxies. The null correlation with halo spin agrees with \citet{Jiang2019}. For the largest, most massive haloes in the `FB-S' galaxies, there may be a slight dependence on the halo spin (bottom right panel). This could be due to the transition from dispersion to rotationally supported galaxies, where the formation of discs becomes increasingly significant \citep{Fall1980,Mo1998}. However, in the MW halo mass regime ($M_{\rm halo} \sim 10^{12}\, \msun$) in FIRE-2 zoom simulations, \citet{Garrison-Kimmel2018} found only weak correlations between the galaxy size and halo spin. Similarly, \citet{Zanisi2020} found in their semi-empirical model using galaxies ($M_\star \sim 10^{9-12}\, \msun$) from the Sloan Digital Sky Survey that the halo spin may not be significant in the GHSR. Including the halo spin in the `FB-S' GHSR does not reduce its scatter.

For the `FB-S' galaxies, the SHMR depends on the halo concentration (top left panel) and spin (top right panel); the GHSR does not depend on either of these halo properties (bottom panels). Thus we suggest that for these low-mass central galaxies, the GHSR is just as or more fundamental than the SHMR, agreeing with semi-empirical results for more massive galaxies \citep{Zanisi2021a,Zanisi2021b}.

\subsection{Galaxy Properties} \label{sec:galprop}

The middle section of Table~\ref{tab:Param} summarises the residual-parameter correlations for the galaxies from all redshifts. Here, the rows in bold represent parameters that decrease the scatter by $>5\%$. Each such power-law index also has $>25\sigma$ difference from 0. 

We define the galaxy shape parameters using only the stellar particles within $\rhalf$. Many of these properties have significant correlations with the scatter. The middle left panel of Figure~\ref{fig:paramplot6panel} displays the figure for the most significant shape property, $\logit[c/a]$. The three axis ratios all have negative correlations, while the elongation, flattening, and triaxiality have positive correlations. Hence, at a fixed halo size, galaxies that are less spherical have larger $\rhalf$. 
This can be expected based on our spherical definition of $\rhalf$. For example, let a triaxial galaxy have axes $a > b > c$ and total stellar mass $M_\star(<a)$. As we integrate the interior stellar mass $M_\star(<r)$ with increasing radius $r$, we reach the edge of the shortest axis first at $r = c$. If $M_\star(<r=c) < 0.5M_\star(<a)$, then we must continue to further radii to determine $\rhalf$. Thus in the direction of the $c$ axis, we do not accumulate any more stellar material. The same argument holds for the $b$ axis, when $M_\star(<r=b) < 0.5M_\star(<a)$. Therefore at a fixed $\rvir$, we expect galaxy sizes to be negatively correlated with the axis ratios and positively correlated with the elongation, flattening, and triaxiality.

Additionally the stellar-to-baryonic mass ratio $\logit[M_\star / M_{\rm bar}]$ -- or similarly the gas-to-baryonic mass ratio $\logit[M_{\rm gas} / M_{\rm bar}] = \logit[1 - M_\star / M_{\rm bar}]$ -- and the stellar-to-total mass ratio $\logit[M_\star / M_{\rm tot}]$ have significant power-law indices and slightly decrease the scatter. These properties are related to the star formation rates and histories of the galaxies, and these properties warrant future study.

We calculate the rotational velocity $v_{\rm rot}$ of the stars,
\begin{equation}
    v_{\rm rot} = \frac{L_\star (<\rhalf)}{M_\star(<\rhalf) \rhalf},
\end{equation}
and the circular velocity $v_{\rm circ}$ at the half-mass radius,
\begin{equation}
    v_{\rm circ} = \sqrt{\frac{G M_{\rm tot}(<\rhalf)}{\rhalf}},
\end{equation}
where $L_\star(\rhalf)$ is the angular momentum of all stars within $\rhalf$. This ratio $v_{\rm rot} / v_{\rm circ}$ determines the amount of rotational support for the galaxy, where $v_{\rm rot} / v_{\rm circ} \sim 1$ is rotationally supported and $\sim 0$ is dispersion supported. The middle right panel of Figure~\ref{fig:paramplot6panel} shows this ratio. We find that at fixed halo size, more discy galaxies (higher $v_{\rm rot} / v_{\rm circ}$) have larger sizes. This amount of rotational support is anti-correlated with $c/a$, meaning that at a fixed $\rvir$, more discy galaxies (lower $c/a$) have more rotational support (higher $v_{\rm rot} / v_{\rm circ}$) and larger sizes. Importantly, most of the low-mass galaxies in our sample are not rotationally supported, with $v_{\rm rot} / v_{\rm circ} \sim 0.0-0.2$. This suggests that our sample is largely dispersion-supported, rather than angular-momentum supported, which helps explain why there is no observed correlation between halo spin and GHSR scatter. However, it remains to be understood why the sizes of low mass galaxies track their halo virial radii so closely.

\subsection{Environment Properties} \label{sec:envirprop}

Lastly, we examine the environmental properties by constructing a comoving sphere of radius $R = 2\, {\rm cMpc}$ and counting each main (central) halo (galaxy) within this sphere (we varied the sphere's radius and found similar results). For each object's catalogue of neighbours, we calculate the distance $\lg{\rm min}(d)$ to the nearest main galaxy; the minimum tidal disruption (Hill) radius $\lg{\rm min}(R_{\rm Hill})$; the mean number $\lg\langle n \rangle$ and mass $\lg \langle \rho \rangle$ (bottom left panel of Figure~\ref{fig:paramplot6panel}) densities and find no significant trends. Lastly we test if galaxies that form spatially near each other systematically lie above or below the GHSR. That is, does galaxy size conformity exist? We correlate the average GHSR offset $\langle \Delta \lg \rhalf \rangle$ of all neighbours (bottom right panel of Figure~\ref{fig:paramplot6panel}), again finding no decrease in scatter for any of the environmental properties.

\section{Conclusions} \label{sec:con}

In this work, we investigate the link between galaxy size and halo size based on a high resolution, cosmological volume simulation suite from the FIRE collaboration. We focus on low-mass centrals with stellar masses $M_\star \sim 10^{7-9}\, \msun$. Our main results and conclusions are as follows:
\begin{itemize}
    \item Our galaxy sizes appear consistent with observations, lying along the extrapolated star forming $M_\star-R_{\rm eff}$ relation at $z=2$ \citep[Figure~\ref{fig:Suess},][]{VanDerWel2014,Mowla2019,Suess2019a,Nedkova2021} and approximating log-normal distributions at $0 \leq z \leq 5$ \citep[Figure~\ref{fig:zall_rhalf_lognormaldist};][]{Shibuya2015}. At $z=0$, the FIREbox galaxies follow the galaxy size-stellar mass relation for star-forming galaxies from \citet{Nedkova2021}, but they are systematically larger by $~\sim 0.3-0.5\, [{\rm dex}]$ at a fixed stellar mass (Figure~\ref{fig:GSSMR}).
    \item The Galaxy-Halo Size Relations (GHSRs) for the low mass objects at each redshift $0 \leq z \leq 5$ are consistent with being linear, agreeing with previous studies \citep{Kravtsov2013,Huang2017,Somerville2018,Jiang2019,Zanisi2020}. The power-law index $\alpha = 0.934\pm0.054$ and scatter $\langle \sigma \rangle = 0.084\, [{\rm dex}]$ from the weighted average from all redshifts are consistent with the individual redshifts, suggesting that the GHSR is constant for low mass galaxies since $z=5$. In general, we find $\rhalf / \rvir \sim 0.05$, which is similar to expectations from spin-based models \citep{Mo1998}.
    \item Whilst the GHSR at each redshift is roughly linear, the power-law index $\alpha_{\rm all-z} = 0.894\pm0.005$ of the combined sample across redshifts suggests that individual objects may trace out paths that are sub-linear (Figure~\ref{fig:zallfits_1024}). This result and the leftward shift of $\lg [ \rhalf / \rvir ]$ distribution with cosmic time (shown in Figure~\ref{fig:lg-rhalf-rvir_probdist_zall_1024}) are qualitatively consistent with expectations of pseudo-evolution of haloes \citep{Diemer2013}. The power-law fit to the all-z sample still details a smooth transition since $z=5$, and the scatter does not decrease when accounting for redshift.
    \item The halo properties we explore -- including spin and concentration -- do {\it not} reduce the scatter in the GHSR in our sample of low-mass, dispersion supported galaxies in FIREbox. Our weak dependence on spin disagrees with classical theoretical ideas of galaxy formation \citep{Mo1998}, but agrees with recent numerical works of more massive galaxies \citep{Desmond2017,Garrison-Kimmel2018,Jiang2019}. At a fixed halo size, the weak correlation of galaxy sizes with halo spin and concentration agree with FIRE-2 zoom simulations \citep{Garrison-Kimmel2018} and semi-empirical models \citep{Zanisi2020,Zanisi2021a} of more massive, rotationally supported galaxies. 
    \item The galaxy shape, amount of stellar rotational support, and potentially the stellar or gas fractions correlate with GHSR residuals. This suggests that baryonic feedback processes of galaxy evolution as well as observable galactic structure/kinematics may be significant in setting the scatter in the GHSR. 
\end{itemize}

The remarkably tight GHSR with nearly constant scatter and linear power-law indices since $z = 5$ allow for estimating halo masses from the sizes of galaxies. This technique is independent from, and as accurate as, other commonly-employed mass-based methods, namely abundance matching, and has the potential to be a better environmental indicator in low-mass galaxy surveys \citep{Yang2007,Yang2009}. Especially with upcoming low-mass galaxy surveys at high redshifts, with, for example, the James Webb Space Telescope, we suggest inferring halo masses and sizes from the galaxy sizes. Given a measurement of a galaxy's stellar mass $M_\star$, one can estimate the halo mass $M_{\rm vir}$ using the SHMR and associated scatter $\sigma \approx 0.25\, [{\rm dex}]$. That is, a halo mass $\lg[M_{\rm vir}(M_\star)/\msun]$ inferred from the stellar mass $M_\star$ has an associated error of roughly $\pm0.25\, [{\rm dex}]$. Now with the galaxy size $\rhalf$, one can estimate the halo size $\rvir$ via the GHSR using the scatter $\sigma \approx 0.08-0.09\, [{\rm dex}]$ for the `FB-S' galaxies. Then a halo size $\lg[\rvir(\rhalf)/\kpc]$ inferred from the galaxy size $R_{1/2}$ has an associated error $\pm0.08-0.09\, [{\rm dex}]$. Because $M_{\rm vir}$ scales exactly as $M_{\rm vir} \propto \rvir^3$, then the error on estimating $\lg[M_{\rm vir}/\msun]$ is three times that of estimating the halo size, given the galaxy size. That is, a halo mass $\lg[M_{\rm vir}(\rhalf)/\msun]$ estimated from the galaxy size $\rhalf$ has an associated error of approximately $\pm0.24-0.27\, {[\rm dex}]$, comparable to the error from using $M_\star$. However, using galaxy sizes could be more beneficial than using stellar masses because the GHSR depends less on the unobservable dark matter halo properties, namely spin and concentration. Thus for dispersion supported, low-mass, central galaxies, we suggest that the galaxy-halo size relation is just as or even more fundamental than the SHMR.

\section*{Acknowledgements}
We extend our deepest appreciation to the innumerable colleagues, staff, and loved ones at each of our institutes and in each others' lives, especially during the COVID-19 pandemic for their continued encouragement and necessary assistance in the research and writing of this paper. 
We thank the anonymous reviewer for their insightful comments, which significantly improved the quality and clarity of this paper. 
We acknowledge PRACE for awarding us access to MareNostrum at the Barcelona Supercomputing Center (BSC), Spain. This research was partly carried out via the Frontera computing project at the Texas Advanced Computing Center. Frontera is made possible by National Science Foundation award OAC-1818253. This work was supported in part by a grant from the Swiss National Supercomputing Centre (CSCS) under project IDs s697 and s698. We acknowledge access to Piz Daint at the Swiss National Supercomputing Centre, Switzerland under the University of Zurich’s share with the project ID uzh18. This work made use of infrastructure services provided by S$^3$IT (\url{www.s3it.uzh.ch}), the Service and Support for Science IT team at the University of Z{\"u}rich.

ECR acknowledges the ThinkSwiss Research Scholarship, funded by the State Secretariat for Education, Research and Innovation (SERI) for the opportunity to spend three months at the University of Z{\"u}rich Institute for Computational Science; travel support from the Alexander Vyssotsky Award from the University of Virginia to present this work; and to many associates -- namely Sven De Rejicke, Arjen van der Wel, Charles Steinhardt, Luca Beale, Robin Leichtnam and Hugues Lascombes -- for their conversations regarding this analysis. ECR is a fellow of the International Max Planck Research School for Astronomy and Cosmic Physics at the University of Heidelberg (IMPRS-HD).

RF acknowledges financial support from the Swiss National Science Foundation (grant no PP00P2\_157591, PP00P2\_194814, and 200021\_188552).

JSB was supported by National Science Foundation (NSF) grant AST-1910346. 

MBK acknowledges support from NSF CAREER award AST-1752913, NSF grant AST-1910346, NASA grant NNX17AG29G, and HST-AR-15006, HST-AR-15809, HST-GO-15658, HST-GO-15901, HST-GO-15902, HST-AR-16159, and HST-GO-16226 from the Space Telescope Science Institute (STScI), which is operated by AURA, Inc., under NASA contract NAS5-26555. 

CAFG was supported by NSF through grants AST-1715216 and CAREER award AST-1652522; by NASA through grant 17-ATP17-0067; by STScI through grant HST-AR-16124.001-A; and by the Research Corporation for Science Advancement through a Cottrell Scholar Award and a Scialog Award.

AW received support from NSF CAREER grant 2045928; NASA ATP grants 80NSSC18K1097 and 80NSSC20K0513; HST grants AR-15057, AR-15809, GO-15902 from STScI; a Scialog Award from the Heising-Simons Foundation; and a Hellman Fellowship.

DK was supported by  NSF grant AST-1715101.

\section*{Data Availability and Software Used}
The data underlying this article will be shared on reasonable request to the corresponding author.

Software used: {\sc Python} \citep{VanDerWalt2011}; {\sc IPython} \citep{Perez2007}; {\sc Numpy} \citep{VanDerWalt2011,Harris2020}; {\sc Scipy} \citep{Virtanen2020}; {\sc Matplotlib} \citep{Hunter2007}; {\sc Jupyter} \citep{Kluyver2016}.

This work made extensive use of the NASA Astrophysics Data System and \url{arXiv.org} preprint server.

\bibliographystyle{mnras}
\bibliography{references}

\appendix

\section{Galaxy Size Definition Checks} \label{app:iter}

\begin{figure*}
    \centering
    \begin{subfigure}[b]{.49\textwidth}
        \centering
        \includegraphics[width=\textwidth]{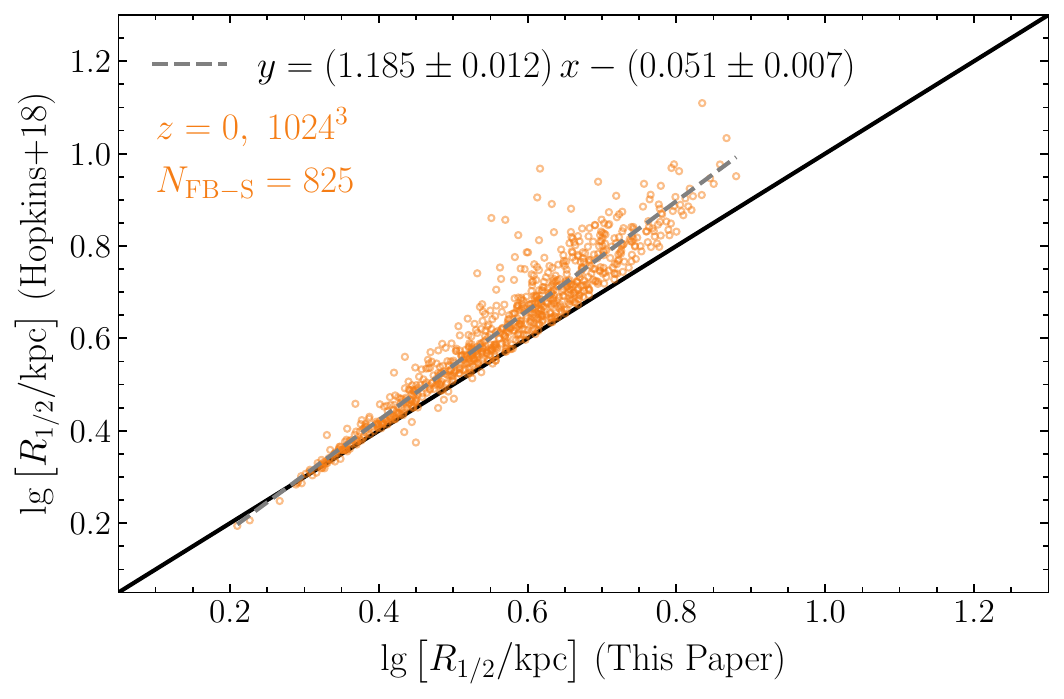}
        \caption{}
        \label{fig:Hopkins_Rhalf_z0}
    \end{subfigure}
    \hfill
    \begin{subfigure}[b]{.49\textwidth}
        \centering
        \includegraphics[width=\textwidth]{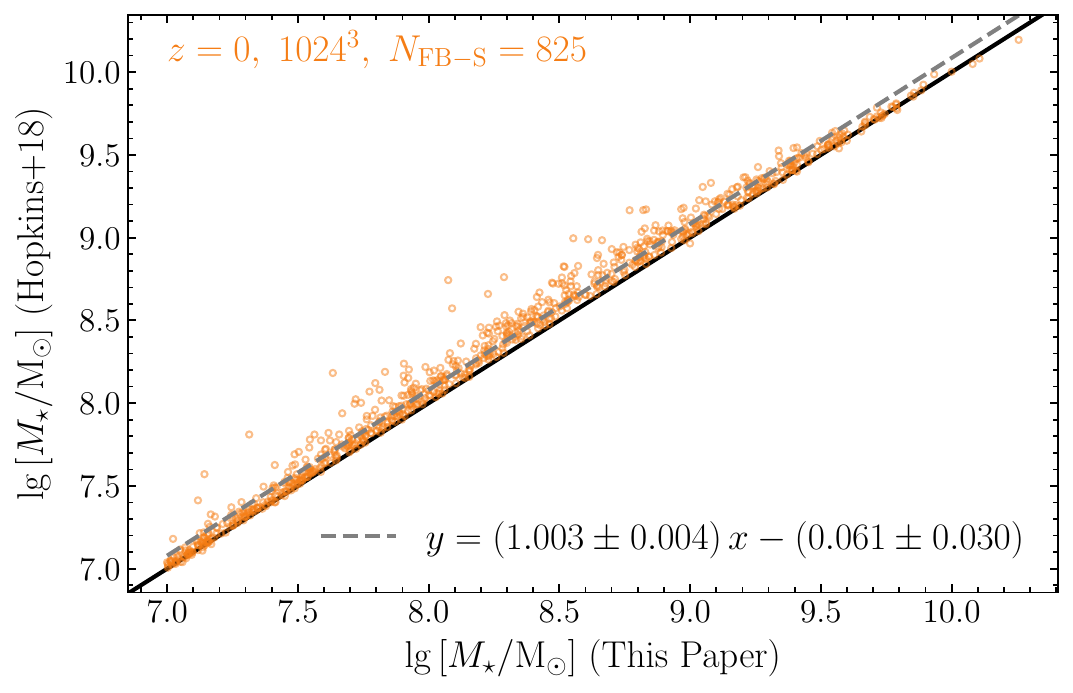}
        \caption{}
        \label{fig:Hopkins_Mstar_z0}
    \end{subfigure}
    \vskip\baselineskip
    \centering
    \begin{subfigure}[b]{.49\textwidth}
        \centering
        \includegraphics[width=\textwidth]{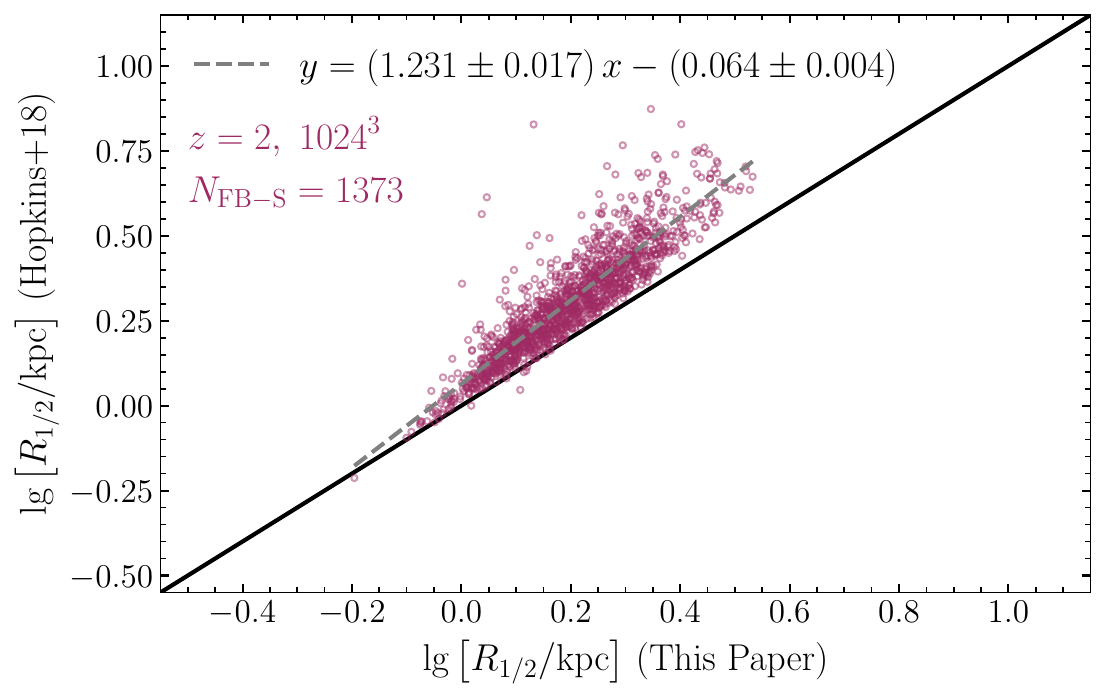}
        \caption{}
        \label{fig:Hopkins_Rhalf_z2}
    \end{subfigure}
    \hfill
    \begin{subfigure}[b]{.49\textwidth}
        \centering
        \includegraphics[width=\textwidth]{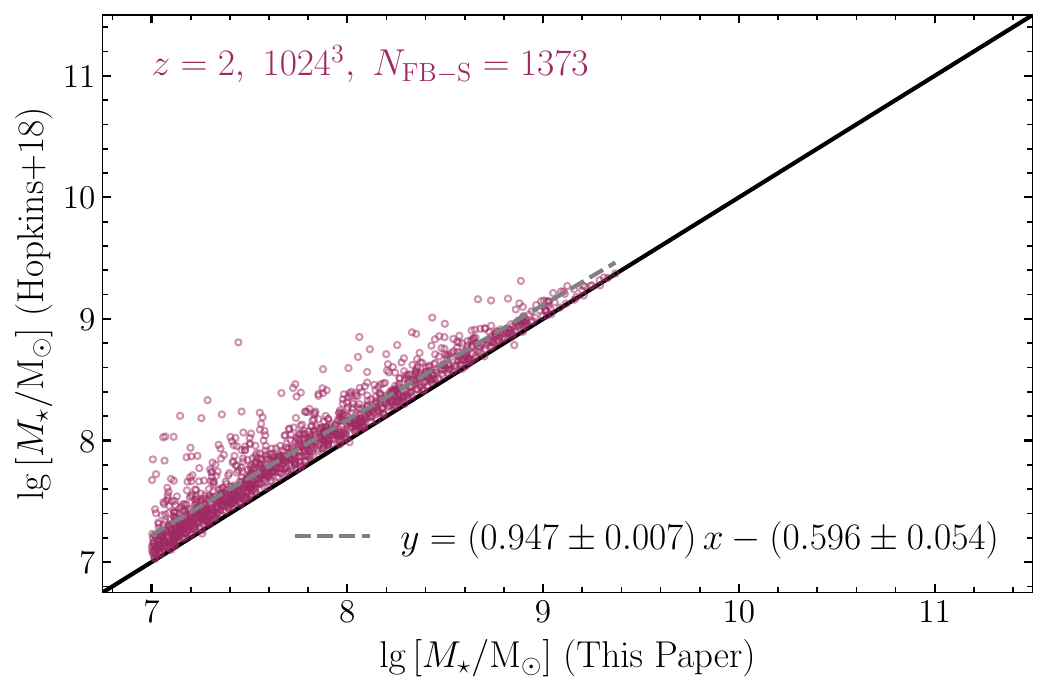}
        \caption{}
        \label{fig:Hopkins_Mstar_z2}
    \end{subfigure}
    \caption{A comparison between galaxy size definitions in the $z=0$ (top) and $z=2$ (bottom) snapshots. The horizontal axes denote the values used in this paper, while the vertical axes the values using the iterative approach \citep{Hopkins2018}. The best fitting power-law equations are given in each panel. The left panels display the galaxy sizes $R_{1/2}$, and the right panels the stellar masses $M_\star$. In general, the values of the galaxy sizes and stellar masses calculated using the two methods agree.}
    \label{fig:Hopkins}
\end{figure*}

\begin{figure}
    \includegraphics[width=\columnwidth]{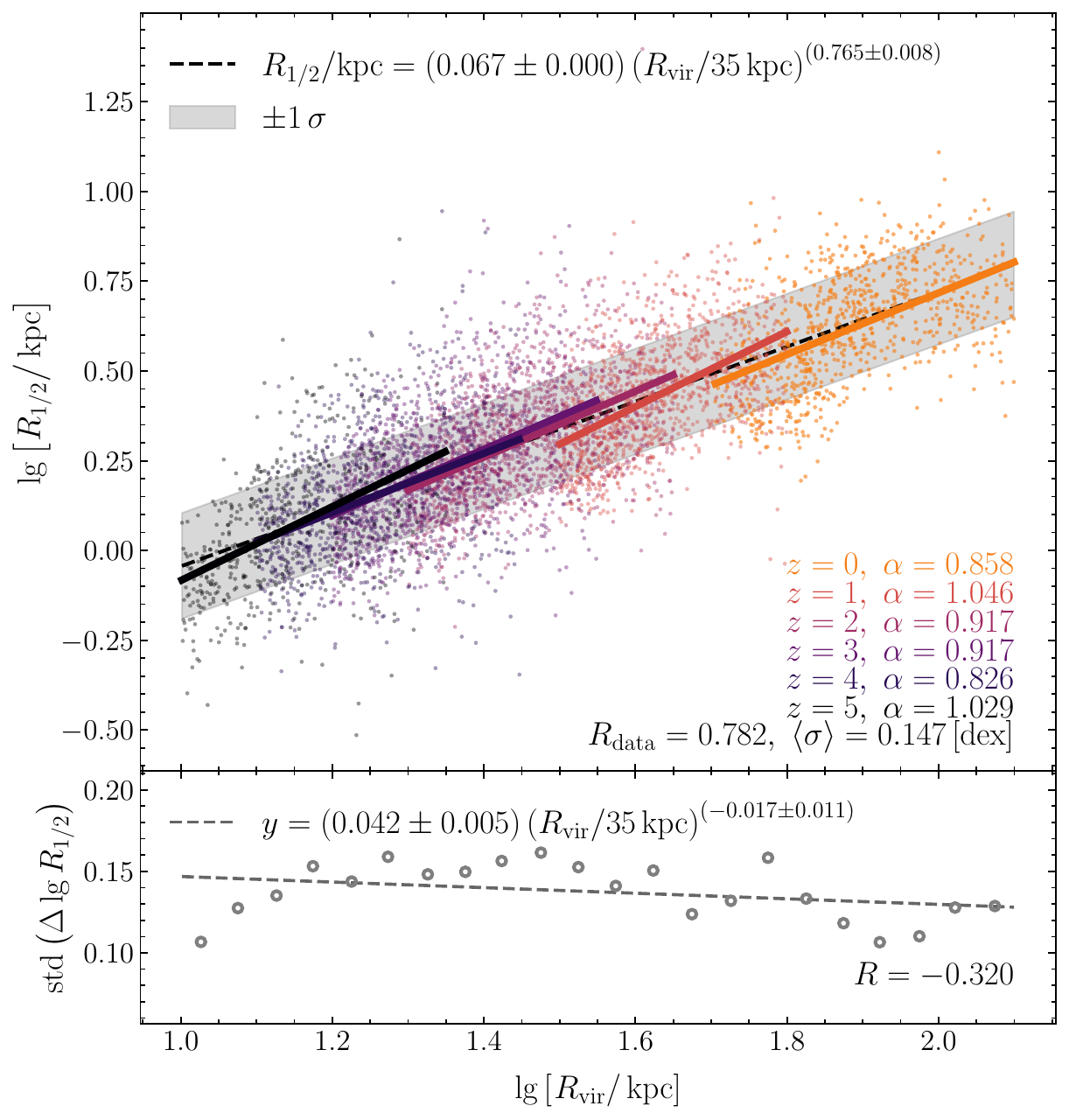}
    \caption{Similar to Figure~\ref{fig:zallfits_1024}, but we use the \citet{Hopkins2018} iterative definition of galaxy size.
    \textit{Top Panel}: the color of the points represents their redshift (see legend), and the best-fitting power-law index $\alpha$ of the GHSR at that redshift is listed on the bottom right. The linear least squares fit to the combined (all$-z$) data is shown at the top of the panel. The Pearson correlation coefficient and the average scatter are listed at the bottom. The power-law relation of the combined data set is sub-linear.
    \textit{Bottom Panel}: The standard deviation of the residuals for each $\lg \rvir$ bin of width $0.050\, [{\rm dex}]$ from the GHSR in the upper panel. The equation of the best-fit and the Pearson correlation coefficient are shown at the top and bottom of the panel. The average scatter $\langle \sigma \rangle = 0.147$ [dex] is larger than that in Figure~\ref{fig:zallfits_1024} $\langle \sigma \rangle = 0.085\, [{\rm dex}]$.}
    \label{fig:zallfits_1024_Hopkins}
\end{figure}

\begin{table*}
 \caption{GHSR at Each and All Redshifts for the Iterative Definition of Galaxy Size from \citet{Hopkins2018} for the FIREbox 1024$^3$ `FB-S' galaxies.}
 \label{tab:GHSR_Hopkins}
 \begin{center}
 \begin{tabular}{cccccc}
  \hline
  $z$ & $\alpha_{\rm GHSR}$ & $\beta_{\rm GHSR}$ & $\langle\sigma\rangle$ & $\alpha_{\sigma}$ & $\beta_\sigma$ \\
  (1) & (2) & (3) & (4) & (5) & (6) \\
  \hline
  $0$ & $0.858\pm0.049$ & $0.061\pm0.003$ & $0.131$ & $-0.020\pm0.055$ & $0.042\pm0.027$ \\
  $1$ & $1.046\pm0.051$ & $0.063\pm0.001$ & $0.134$ & $0.134\pm0.051$ & $0.023\pm0.022$ \\
  $2$ & $0.917\pm0.043$ & $0.070\pm0.001$ & $0.137$ & $0.259\pm0.045$ & $0.016\pm0.004$\\
  $3$ & $0.917\pm0.050$ & $0.074\pm0.002$ & $0.151$ & $0.368\pm0.023$ & $0.012\pm0.001$ \\
  $4$ & $0.826\pm0.068$ & $0.070\pm0.004$ & $0.166$ & $0.473\pm0.035$ & $0.010\pm0.001$\\
  $5$ & $1.029\pm0.093$ & $0.086\pm0.007$ & $0.174$ & $0.461\pm0.041$ & $0.013\pm0.002$ \\
  \hline
  mean$^a$ & $0.930\pm0.076$ & $0.070\pm0.018$ & $0.146$ & $0.26\pm0.20$ & $0.013\pm0.011$ \\
  all$-z^b$ & $0.765\pm0.008$ & $0.067\pm0.001$ & $0.147$ & $-0.017\pm0.011$ & $0.042\pm0.005$ \\
  \hline
 \end{tabular}
\end{center}
 \parbox{\textwidth}{
  \footnotesize{
  $^a$The averages using the number of objects at each redshift as weights. \\ 
  $^b$The combined sample of all redshift snapshots, treating objects from different snapshots equally. \\
  The same as Table~\ref{tab:GHSR}, here using the iterative definition of galaxy size \citep{Hopkins2018}. (1) The redshift; (2) and (3) the power-law index and normalisation for the GHSR from Equation~\eqref{eqn:pl}; (4) the average scatter in the GHSR; (5) and (6) the power-law index and normalisation for the GHSR scatter as a function of $\lg\rvir$ bin.
  }
 }
\end{table*}

\begin{figure*}
    \centering
    \begin{subfigure}[b]{.49\textwidth}
        \centering
        \includegraphics[width=\textwidth]{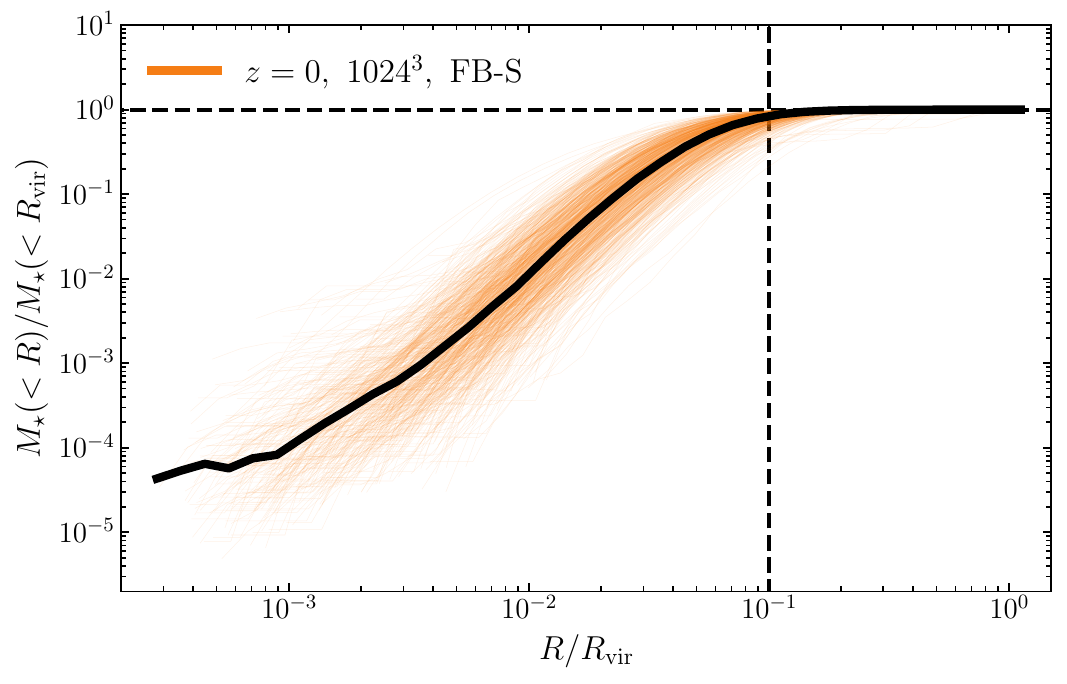}
        \label{fig:mstar-radprof_z0}
    \end{subfigure}
    \hfill
    \begin{subfigure}[b]{.49\textwidth}
        \centering
        \includegraphics[width=\textwidth]{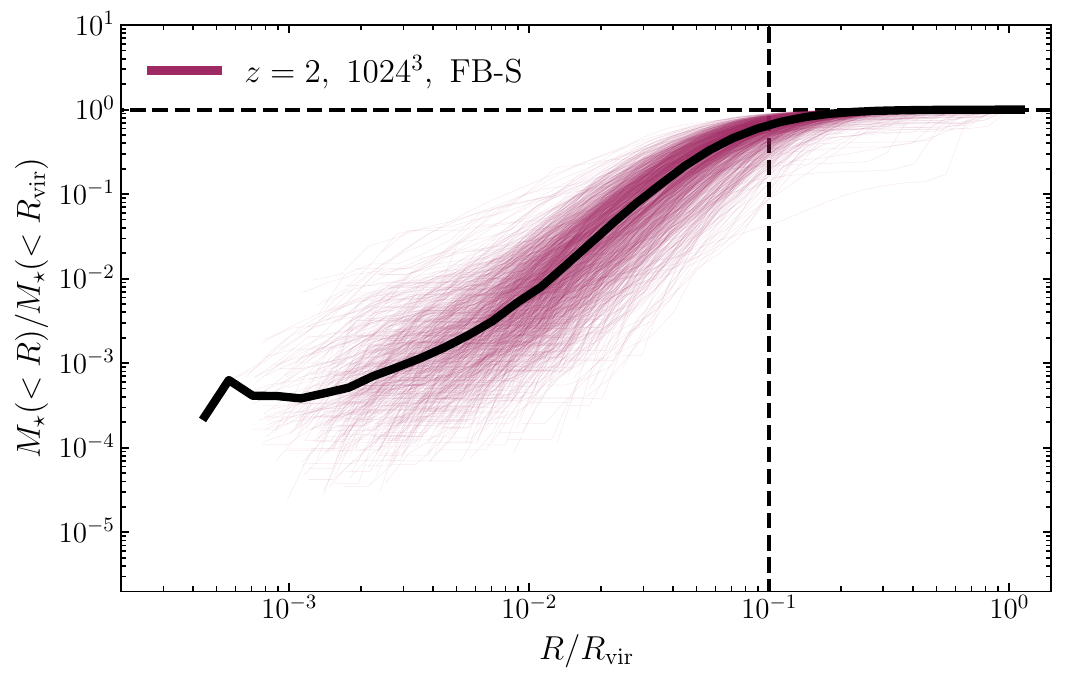}
        \label{fig:mstar-radprof_z2}
    \end{subfigure}
    \caption{The stellar mass profiles for the galaxies in FIREbox pathfinder at redshift $z=0$ (left) and $z=2$ (right). The thick black curves show the median values within $R/\rvir$ bins of width $0.1\, [{\rm dex}]$. We normalize the stellar masses $M_\star(<R)$ by the total stellar mass within the halo $M_\star(<\rvir)$, and the radius $R$ by virial radius $\rvir$. We mark where $M_\star(<R) = M_\star(<\rvir)$ and $R = 0.1\rvir$ with horizontal and vertical dashed lines, respectively. The median curves flatten to $M_\star \approx M_\star(<\rvir)$ by $0.1\rvir$, affirming that this cutoff radius contains most of the haloes' stellar mass.}
    \label{fig:mstar_radprof}
\end{figure*}

We check our definition of galaxy size -- $M_\star(<\rhalf) = 0.5 M_\star(<0.1\rvir)$ -- versus an iterative approach. \citet{Hopkins2018} calculates an initial three-dimensional half-stellar mass radius $R_{1/2,0}$ within a large cutoff at $0.15\rvir$. Then they define an intermediate total radius $R_{\rm tot} = 3R_{1/2,0}$ and recalculate the half-stellar mass radius: $M_\star(<R_{1/2,1}) = 0.5M_\star(<R_{\rm tot}) = 0.5M_\star(<3R_{1/2,0})$. They repeat this process until convergence
\begin{equation}
    \frac{\lvert R_{1/2,n} - R_{1/2,n+1} \rvert}{R_{1/2,n+1}} < \epsilon
\end{equation}
for some tolerance $\epsilon$, where $n$ is the number of iterations. We choose $\epsilon = 10^{-5}$, which typically takes $n \lesssim 10$ iterations. We require convergence within $n_{\rm max} = 100$ iterations, within which every galaxy size estimate converges.  

Figure~\ref{fig:Hopkins} details the similarities between these definitions, where the left panels compare the galaxy sizes $\rhalf$ directly and the right panels the stellar masses $M_\star$ at $z=0$ (top) and $z=2$ (bottom). In general, there is good agreement between the definitions, except at the smallest and largest galaxy sizes. \citet{Hopkins2018} warns at lower redshifts our definition fails because satellites can exist within $0.1\rvir$, but this typically affects more massive galaxies. We also check our results using all stellar material within $0.2\rvir$ and find qualitatively similar results. The top panels at $z=0$ show good agreement between the two galaxy size definitions. 

We repeat Table~\ref{tab:GHSR} and Figure~\ref{fig:zallfits_1024} using the iterative galaxy definition in Table~\ref{tab:GHSR_Hopkins} and Figure~\ref{fig:zallfits_1024_Hopkins}. The same qualitative results hold, except there is larger overall scatter $\langle\sigma\rangle = 0.146\, [{\rm dex}]$ compared to that in Figure~\ref{fig:zallfits_1024} ($\langle \sigma \rangle = 0.085\, [{\rm dex}]$). At any given instant in time the GHSR is roughly linear, but the size of a given halo typically grows faster than the size of its central galaxy over much of cosmic history.

As a final check for our definition of $\rhalf$, we plot the stellar mass radial profiles in Figure~\ref{fig:mstar_radprof} for the galaxies in the $1024^3$ run at redshift $z=0$ (left) and $z-2$ (right). The median stellar mass curves flatten to $M_\star \approx M_\star(<\rvir)$ (horizontal dashed line) by the total galaxy size $R \approx 0.1\rvir$ (vertical dashed line), affirming that this is a robust value of the total galaxy size.

\section{Galaxy Size-Stellar Mass Relations} \label{app:GSSMR}

\begin{figure*}
    \centering
    \begin{subfigure}[b]{.49\textwidth}
        \centering
        \includegraphics[width=\textwidth]{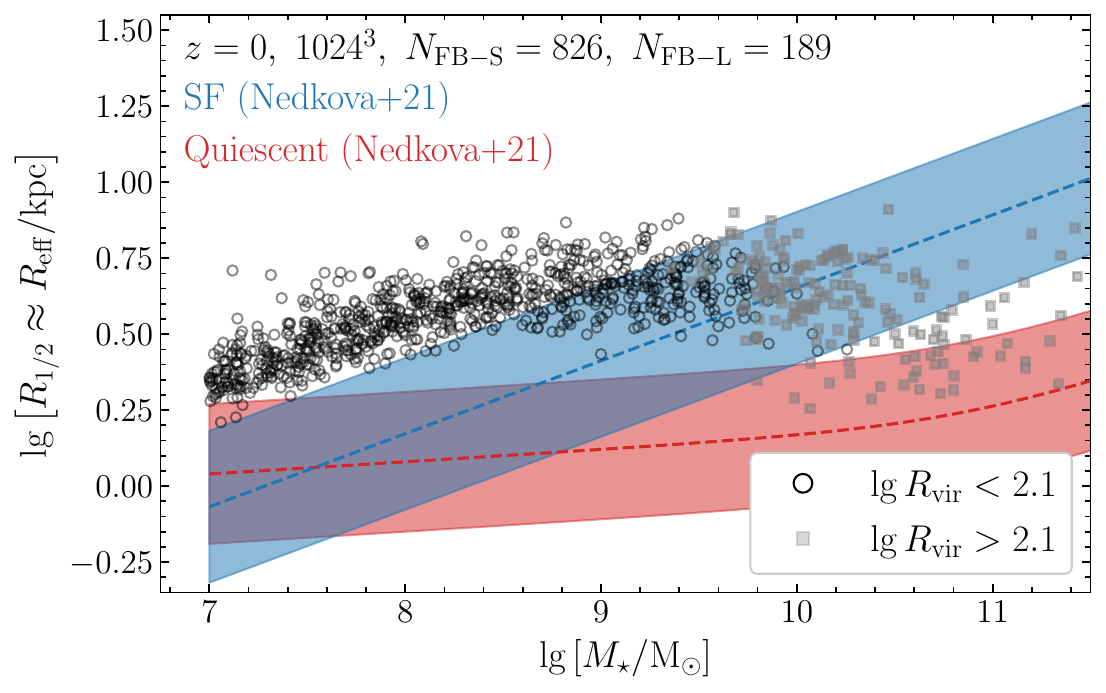}
        \label{fig:rhalfmstar_z0}
    \end{subfigure}
    \hfill
    \begin{subfigure}[b]{.49\textwidth}
        \centering
        \includegraphics[width=\textwidth]{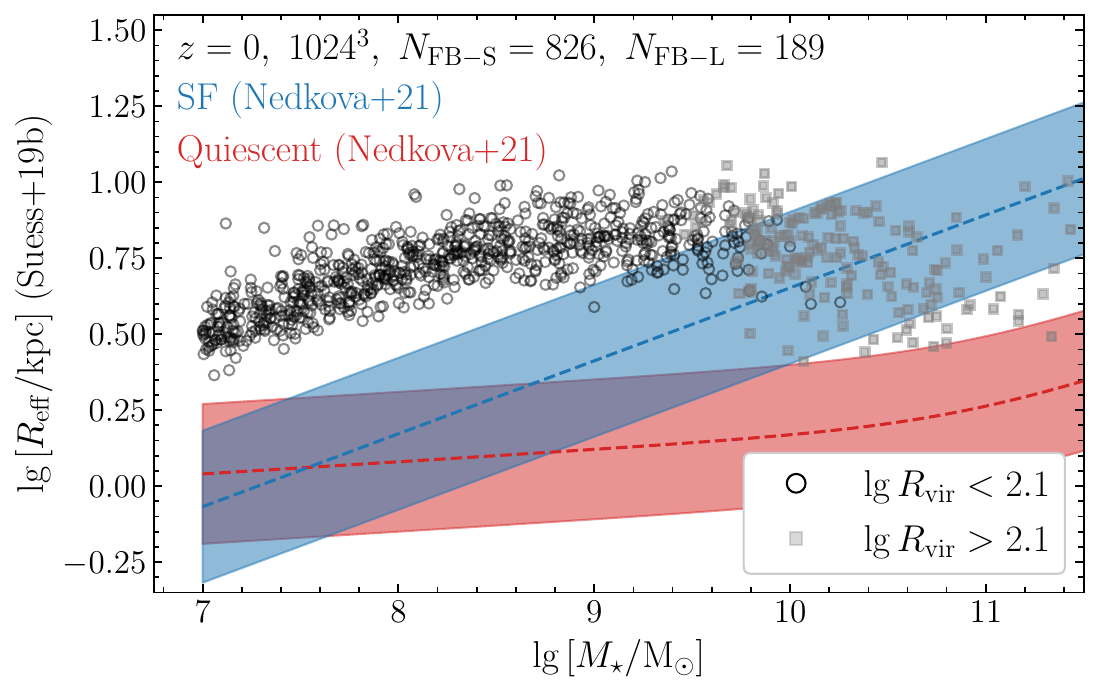}
        \label{fig:mreffmstar_z2}
    \end{subfigure}
    \caption{Similar to Figure~\ref{fig:Suess}. A comparison to the galaxy size-stellar mass relation obtained from the Hubble Frontier Fields (HFF) and Cosmic Assembly Near-infrared Deep Extragalactic Survey (CANDELS) at $z=0$ \citep{Nedkova2021}. In the left panel, we approximate the effective radius $R_{\rm eff}$ from \citet{Nedkova2021} with our half-mass radius $\rhalf$; in the right panel, we convert $\rhalf$ to $R_{\rm eff}$ using the correction factor from \citet{Suess2019b}. Specifically, \citet{Suess2019b} find for both quiescent and star-forming galaxies of stellar mass $M_\star > 10^{10.1}\, \msun$ at redshift $z \lesssim 1$ that the 2D half-mass radius $R_{\rm mass} \sim 0.7R_{\rm eff}$. The `FB-S' galaxies follow a similar galaxy size-stellar mass relation to the star-forming galaxies from \citet{Nedkova2021}, but the `FB-S' galaxies are systematically larger by $\sim0.5\, [{\rm dex}]$.}
    \label{fig:GSSMR}
\end{figure*}

In Figure~\ref{fig:Suess}, we use a correction from \citet{Suess2019a} to convert our 3D intrinsic half-mass radii $\rhalf$ to 2D projected effective half-light radii $R_{\rm eff}$ at $z=2$, assuming that the 3D $\rhalf$ has on average a similar value to the 2D half-mass radius $R_{\rm mass}$ \citep{Ven2021}. The completeness limit for \citet{Suess2019a} is $M_\star \gtrsim 10^{10}\, \msun$, and it is unclear if this holds in the stellar mass regime of the `FB-S' galaxies ($M_\star \sim 10^{7-9}\, \msun$). However, the correction does not significantly affect the results at $z=2$, where $\langle \rhalf / R_{\rm eff} \rangle \sim 0.9$ for both star-forming and quiescent galaxies. Without this correction, the `FB-S' galaxies are still within the scatter of the star-forming relation.

In Figure~\ref{fig:GSSMR}, we compare our galaxy size-stellar mass relations to those observed at $z=0$ \citep{Nedkova2021} in the Hubble Frontier Fields (HFF) and Cosmic Assembly Near-infrared Deep Extragalactic Survey (CANDELS). In the left panel, we approximate the effective radius $R_{\rm eff}$ as our half-mass radius $\rhalf$; in the right panel, we convert $\rhalf$ to $R_{\rm eff}$ using the correction factor from \citet{Suess2019b}, assuming $\rhalf \sim R_{\rm mass}$. Specifically, \citet{Suess2019b} find for both quiescent and star-forming galaxies of stellar mass $M_\star > 10^{10.1}\, \msun$ at redshift $z \lesssim 1$ that $\langle R_{\rm mass} / R_{\rm eff} \rangle \sim 0.7$. It is unclear if this correction holds at the stellar masses of the `FB-S' galaxies ($M_\star \sim 10^{7-9.5}\, \msun$) at $z=0$. At a fixed stellar mass, the `FB-S' galaxies systematically have a half-mass radius $\rhalf$ larger than the star-forming galaxies from \citet{Nedkova2021}. At stellar masses $M_\star \sim 10^{9-10}\, \msun$, the `FB-S' and `FB-L' galaxy sizes flatten out. Then the `FB-L' galaxy sizes turnover at $M_\star \sim 10^{10}\, \msun$, where $\rhalf$ starts to decrease with increasing stellar mass. Without the correction from \citet[][left panel]{Suess2019b}, the `FB-S' galaxy sizes are systematically $\sim 0.3\, [{\rm dex}]$ larger than observations. Then the `FB-L' galaxies lie partly between the galaxy size-stellar mass relations for star-forming and quiescent galaxies. With the correction (right panel), the galaxy size offset is $\sim 0.5\, [{\rm dex}]$, and the `FB-L' galaxies are consistent with the size-mass relation for star-forming galaxies.

\section{Additional Halo Properties and the GHSR Scatter} \label{app:dmo}

\begin{figure*}
    \includegraphics[width=\textwidth]{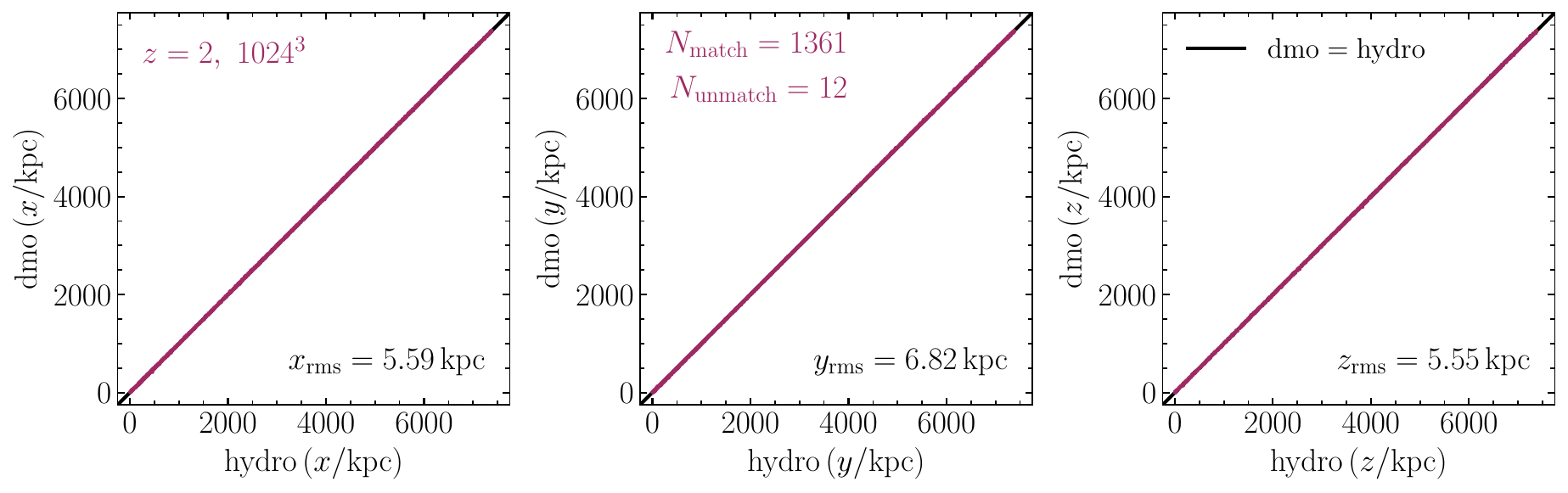}
    \caption{A comparison between the halo centers in the hydrodynamical (horizontal axis) and DMO simulation (vertical axis). There is strong agreement in the halo centers (shown above) and the masses (not shown).}
    \label{fig:dmo}
\end{figure*}

Similarly to Table~\ref{tab:Param}, Table~\ref{tab:Param_alt} summarises the correlation fits between the GHSR scatter and additional halo properties at $z=2$.

The top set of parameters correlates the GHSR residuals with the difference between the property of a given halo and the mean property of all haloes within the $\lg \rvir$ bin of width $0.050\, [{\rm dex}]$. Similarly to the direct value of the halo property, only the center-of-mass offset has a slope significantly different from 0, but no offset parameters decrease the scatter by $>5\%$.

So far, the halo properties are always from the hydrodynamic simulation, but perhaps the baryonic properties within the halo affect the dark matter properties, specifically the spin and concentration. We run FIREbox with dark matter only (DMO) and with all particles (ALL-hydro), and we create a mapping between haloes in the hydro and DMO simulations. 

First, we create two AHF catalogues from the hydro simulation: one using only dark matter particles (DMO-hydro), one using all particles (ALL). We correlate haloes between these catalogues using the spatial coordinates of the halo centers. Specifically, we create a sphere of radius $\rvir$ centered on the ALL-halo center's coordinates in the DMO-hydro catalogue and tabulate every DMO-hydro halo within this sphere. Then we choose the halo that is most similar in mass to the ALL-halo as the best match. This completes the mapping between the two AHF catalogues generated from the hydro simulation.

Next, we map the AHF catalogue from the DMO simulation with that of the DMO-hydro using the dark matter particle IDs. We use AHF \textsc{mergertree} to link DMO-hydro haloes to those of the DMO. The likelihood of connection is
\begin{equation}
    \mathrm{max_j \left( M_{ij} = \frac{N^2_{i \cap j}}{N_i N_j} \right)},
    \label{fig:AHF mergertree merit function}
\end{equation}
where ${\rm M_{ij}}$ is the merit function, ${\rm N_{i \cap j}}$ is the number of shared particles between a DMO and its corresponding DMO-hydro halo, while ${\rm N_{i}}$ and ${\rm N_{j}}$ are the total number of particles in the DMO and DMO-hydro haloes respectively. This completes the mapping between the DMO-hydro and DMO halo catalogues.

Lastly, we combine the two maps to correlate the ALL-hydro and DMO haloes. We require that the DMO halo is a main halo, since the ALL-hydro halo is by a main halo by definition. Then we require the DMO virial mass be within a factor of 2 of that of the ALL-hydro halo. There are no explicit requirements on the halo centers between these catalogues, and we use this as a final check of accurate mapping. Figure~\ref{fig:dmo} details the positions, and we find that the halo centers are in great agreement. We successfully map 1361/1373 haloes at $z-2$, where each halo is central, and has similar size, mass and coordinates. The same results hold at the other redshifts.

Thus, we correlate the GHSR residuals with the pristine dark matter properties of the matched haloes in the DMO simulation (middle set of Table~\ref{tab:Param_alt}). Nevertheless, the DMO halo properties do not decrease the GHSR scatter, and the COM offset has a weaker significance. We expect that the lack of satellite galaxies causes the typical COM values to be smaller, causing the weaker correlation. 

The bottom set combines the the techniques of the second and third sets of correlations by correlating the GHSR residuals with the difference between the DMO property of a given halo and the mean property of all DMO haloes in the same $\rvir$ bin. Agreeing with our previous results, we find that no studied halo property significantly explains the scatter in the GHSR in the `FB-S' galaxies since $z=5$.

\begin{table*}
 \caption{Correlations of the `FB-S' GHSR's Scatter with Additional Halo Properties at all redshifts.}
 \label{tab:Param_alt}
 \begin{center}
 \begin{tabular}{lcccccc}
  \hline
  Parameter & $m$ & $\sigma_{m,0}$ & $b$ & $R$ & $\langle\sigma\rangle$  &  $\%\Delta\sigma$ \\
  (1) & (2) & (3) & (4) & (5) & (6) & (7) \\
  \hline
$\Delta {\rm logit}\left[ b/a\ \left({\rm halo}, \rvir\right) \right]$ & $0.016\pm0.003$ & $4.76$ & $-0.001\pm0.001$ & $0.055$ & $0.085\pm0.001$ & $0.2$ \\
$\Delta {\rm logit}\left[ c/a\ \left({\rm halo}, \rvir\right) \right]$ & $0.027\pm0.005$ & $5.17$ & $-0.001\pm0.001$ & $0.058$ & $0.085\pm0.001$ & $0.2$ \\
$\Delta {\rm logit}\left[ c/b\ \left({\rm halo}, \rvir\right) \right]$ & $0.004\pm0.004$ & $1.09$ & $-0.000\pm0.001$ & $0.013$ & $0.085\pm0.001$ & $0.0$ \\
$ \Delta{\rm logit}\left[ E \equiv \sqrt{1 - (b/a)^2}\ \left({\rm halo}, \rvir\right) \right]$ & $-0.021\pm0.004$ & $4.99$ & $-0.001\pm0.001$ & $-0.058$ & $0.085\pm0.001$ & $0.2$ \\
$ \Delta{\rm logit}\left[ F \equiv \sqrt{1-(c/b)^2}\ \left({\rm halo}, \rvir\right) \right]$ & $-0.005\pm0.005$ & $1.08$ & $-0.000\pm0.001$ & $-0.013$ & $0.085\pm0.001$ & $0.0$ \\
$ \Delta{\rm logit}\left[ T \equiv \left.\left(1 - \left(b/a\right)^2\right)\right/\left(1- \left(c/a\right)^2\right)\ \left({\rm halo}, \rvir\right) \right]$ & $-0.007\pm0.002$ & $3.08$ & $-0.000\pm0.001$ & $-0.037$ & $0.085\pm0.001$ & $0.1$ \\
$\Delta \lg \lambda$ & $0.030\pm0.004$ & $6.93$ & $0.002\pm0.001$ & $0.103$ & $0.084\pm0.001$ & $0.4$ \\
$\Delta \lg \lambda_e$ & $0.028\pm0.005$ & $6.11$ & $0.002\pm0.001$ & $0.086$ & $0.084\pm0.001$ & $0.3$ \\
$\Delta \lg {\rm cNFW}$ & $0.000\pm0.008$ & $0.06$ & $0.000\pm0.001$ & $-0.013$ & $0.085\pm0.001$ & $0.0$ \\
$\Delta \lg\left[\sigma_v/({\rm km\, s}^{-1})\right]$ & $0.010\pm0.011$ & $0.99$ & $0.000\pm0.001$ & $0.030$ & $0.085\pm0.001$ & $0.0$ \\
$\Delta \lg\left[\Delta {\rm COM}/{\rm kpc}\right]$ & $0.063\pm0.004$ & $18.02$ & $0.007\pm0.001$ & $0.247$ & $0.082\pm0.001$ & $2.7$ \\
\hline
${\rm dmo}\left({\rm logit}\left[ b/a\ ({\rm halo}, \rvir) \right]\right)$ & $0.011\pm0.003$ & $3.32$ & $-0.009\pm0.003$ & $0.044$ & $0.084\pm0.001$ & $0.1$ \\
${\rm dmo}\left({\rm logit}\left[ c/a\ ({\rm halo}, \rvir) \right]\right)$ & $0.017\pm0.005$ & $3.54$ & $-0.009\pm0.003$ & $0.047$ & $0.084\pm0.001$ & $0.1$ \\
${\rm dmo}\left({\rm logit}\left[ c/b\ \left({\rm halo}, \rvir\right) \right]\right)$ & $0.003\pm0.004$ & $0.76$ & $-0.003\pm0.004$ & $0.010$ & $0.084\pm0.001$ & $0.0$ \\
${\rm dmo}\left({\rm logit}\left[ E \equiv \sqrt{1 - (b/a)^2}\ \left({\rm halo}, \rvir\right) \right]\right)$ & $-0.015\pm0.004$ & $3.58$ & $-0.000\pm0.001$ & $-0.048$ & $0.084\pm0.001$ & $0.1$ \\
${\rm dmo}\left({\rm logit}\left[ F \equiv \sqrt{1-(c/b)^2}\ \left({\rm halo}, \rvir\right) \right]\right)$ & $-0.004\pm0.005$ & $0.78$ & $-0.000\pm0.001$ & $-0.010$ & $0.084\pm0.001$ & $0.0$ \\
${\rm dmo}\left({\rm logit}\left[ T \equiv \left.\left(1 - \left(b/a\right)^2\right)\right/\left(1- \left(c/a\right)^2\right)\ \left({\rm halo}, \rvir\right) \right]\right)$ & $-0.006\pm0.002$ & $2.32$ & $0.001\pm0.001$ & $-0.031$ & $0.084\pm0.001$ & $0.0$ \\
${\rm dmo}\left(\lg\lambda\right)$ & $0.022\pm0.009$ & $2.39$ & $0.029\pm0.016$ & $0.048$ & $0.084\pm0.002$ & $0.1$ \\
${\rm dmo}\left(\lg\lambda_e\right)$ & $0.022\pm0.008$ & $2.70$ & $0.032\pm0.014$ & $0.049$ & $0.085\pm0.002$ & $0.1$ \\
${\rm dmo}\left(\lg{\rm cNFW}\right)$ & $-0.018\pm0.004$ & $4.02$ & $0.014\pm0.004$ & $-0.054$ & $0.084\pm0.001$ & $0.1$ \\
${\rm dmo}\left(\lg\left[\sigma_{v}/({\rm km\, s}^{-1})\right]\right)$ & $0.001\pm0.010$ & $0.09$ & $-0.002\pm0.018$ & $0.001$ & $0.084\pm0.001$ & $0.0$ \\
${\rm dmo}\left(\lg\left[\Delta{\rm COM}/{\rm kpc}\right]\right)$ & $0.009\pm0.003$ & $2.98$ & $-0.005\pm0.002$ & $0.040$ & $0.084\pm0.001$ & $0.1$ \\
\hline
${\rm dmo} \left( \Delta{\rm logit}\left[ b/a\ \left({\rm halo}, \rvir\right) \right]\right)$ & $0.013\pm0.003$ & $3.85$ & $-0.001\pm0.001$ & $0.044$ & $0.084\pm0.001$ & $0.1$ \\
${\rm dmo} \left( \Delta{\rm logit}\left[ c/a\ \left({\rm halo}, \rvir\right) \right]\right)$ & $0.023\pm0.005$ & $4.32$ & $-0.001\pm0.001$ & $0.047$ & $0.084\pm0.001$ & $0.2$ \\
${\rm dmo} \left( \Delta{\rm logit}\left[ c/b\ \left({\rm halo}, \rvir\right) \right]\right)$ & $0.004\pm0.004$ & $1.03$ & $-0.000\pm0.001$ & $0.010$ & $0.084\pm0.001$ & $0.0$ \\
${\rm dmo} \left( \Delta{\rm logit}\left[ E \equiv \sqrt{1 - (b/a)^2}\ \left({\rm halo}, \rvir\right) \right]\right)$ & $-0.018\pm0.004$ & $4.12$ & $-0.001\pm0.001$ & $-0.048$ & $0.084\pm0.001$ & $0.2$ \\
${\rm dmo} \left( \Delta{\rm logit}\left[ F \equiv \sqrt{1-(c/b)^2}\ \left({\rm halo}, \rvir\right) \right]\right)$ & $-0.005\pm0.005$ & $1.05$ & $-0.000\pm0.001$ & $-0.010$ & $0.084\pm0.001$ & $0.0$ \\
${\rm dmo} \left( \Delta{\rm logit}\left[ T \equiv \left.\left(1 - \left(b/a\right)^2\right)\right/\left(1- \left(c/a\right)^2\right)\ \left({\rm halo}, \rvir\right) \right]\right)$ & $-0.006\pm0.002$ & $2.49$ & $-0.000\pm0.001$ & $-0.031$ & $0.084\pm0.001$ & $0.1$ \\
${\rm dmo}\left(\Delta \lg \lambda\right)$ & $0.017\pm0.009$ & $1.89$ & $-0.003\pm0.003$ & $0.048$ & $0.084\pm0.002$ & $0.1$ \\
${\rm dmo}\left(\Delta \lg \lambda_e\right)$ & $0.020\pm0.008$ & $2.44$ & $-0.001\pm0.002$ & $0.049$ & $0.085\pm0.002$ & $0.1$ \\
${\rm  dmo}\left(\Delta \lg {\rm cNFW}\right)$ & $-0.030\pm0.007$ & $4.48$ & $-0.001\pm0.001$ & $-0.054$ & $0.084\pm0.001$ & $0.2$ \\
${\rm dmo}\left(\Delta \lg\left[\sigma_v/({\rm km\, s}^{-1})\right]\right)$ & $-0.015\pm0.011$ & $1.34$ & $-0.000\pm0.001$ & $0.001$ & $0.084\pm0.001$ & $0.0$ \\
${\rm dmo}\left(\Delta \lg\left[\Delta {\rm COM}/{\rm kpc}\right]\right)$ & $0.005\pm0.003$ & $1.68$ & $0.001\pm0.001$ & $0.040$ & $0.084\pm0.001$ & $0.0$ \\
\hline
\end{tabular}
 \end{center}
 \parbox{\textwidth}{
  \footnotesize{
  The same as Table~\ref{tab:Param}, except only for the alternative halo properties.
  (1) The parameter used as the horizontal axis; (2) the slope of the fit; (3) the $\sigma$ from the slope being $0$ using the statistical error of $m$; (4) the vertical offset in the fit; (5) the Pearson correlation coefficient; (6) the scatter in regression in the residuals versus parameter; (7) the percentage difference in scatter between the GHSR and the residual-parameter relation. From top to bottom the sections are the difference between the value and the average value of similar-massed haloes, the value of the cross-matched halo in the dark matter only (DMO) simulation, and the difference between the cross-matched halo value and the average value of all similar-massed DMO haloes. \S~\ref{sec:haloprop} describes what each parameter is.
  }
 }
\end{table*}

\bsp	
\label{lastpage}
\end{document}